\documentclass[prb,superscriptaddress,twocolumn,floatfix,notitlepage,citeautoscript]{revtex4-2}
\usepackage[T2A]{fontenc}
\usepackage{amsmath}
\usepackage{graphicx}
\usepackage{color}
\usepackage{epstopdf}
\begin{document}
\title{Finite-temperature Hatree--Fock--Bogoliubov theory for exciton-polaritons}

\author{A.~M.~Grudinina}
\affiliation{National Research Nuclear University MEPhI (Moscow Engineering Physics Institute), 115409 Moscow, Russia}

\author{I.~L.~Kurbakov}
\affiliation{Institute for Spectroscopy, Russian Academy of Sciences, 142190 Troitsk, Moscow, Russia}

\author{Yu.~E.~Lozovik}
\affiliation{Institute for Spectroscopy, Russian Academy of Sciences, 142190 Troitsk, Moscow, Russia}
\affiliation{MIEM, National Research University Higher School of Economics, 101000 Moscow, Russia}

\author{N.~S.~Voronova}
\email{nsvoronova@mephi.ru}
\affiliation{National Research Nuclear University MEPhI (Moscow Engineering Physics Institute), 115409 Moscow, Russia}
\affiliation{Russian Quantum Center, Skolkovo IC, Bolshoy boulevard 30 bld. 1, 121205 Moscow, Russia}

\begin{abstract}
Microcavity exciton-polaritons, known to exhibit non-equilibrium Bose condensation at high critical temperatures, can be also brought in thermal equilibrium with the surrounding medium and form a quantum degenerate Bose-Einstein distribution. It happens when their thermalization time in the regime of positive detunings---or, alternatively, for high-finesse microcavities---becomes shorter than their lifetime. Here we present the self-consistent finite-temperature Hartree--Fock--Bogoliubov description for such a system of polaritons, universally addressing the excitation spectrum, momentum-dependent interactions, condensate depletion, and the background population of dark excitons that contribute to the system's chemical potential. Employing the derived expressions, we discuss the implications for the Bogoliubov sound velocity, confirmed by existing experiments, and  define the critical temperatures of (quasi-)condensation and the integral particle lifetime dependencies on the detuning. Large positive detunings are shown to provide conditions for the total lifetime reaching nanosecond timescales. This allows realization of \textit{thermodynamically-equilibrium} polariton systems with Bose-Einstein condensate forming at temperatures as high as tens of Kelvin.
\end{abstract}

\maketitle

\section{Introduction}

Spectra of low-lying collective excitations in weakly interacting Bose gases, essential to describe superfluidity in quantum fluids, are strongly dependent on correlations induced in the Bose condensate and the shape of both the single-particle spectrum and the interaction potential.
Additionally, in the systems where the Galilean invariance is broken, like in systems with momentum-dependent mass and particle interactions, the Landau criterion for superfluidity is inapplicable and there is no way to define the superfluid density through the regular calculation of the mass flow \cite{keeling2006,semenov}. One recent example of such a system is a Bose gas of exciton-polaritons---half light, half matter two-dimensional (2D) quasiparticles---whose macroscopic degeneracy is routinely observed at high critical temperatures nowadays \cite{Microcavities,RMP2010,QFL}.

When discussing Bose condensation, the lower branch of the exciton-polariton energy dispersion becomes of interest. It has an essentially non-parabolic shape strongly dependent on the detuning between the cavity photon and the exciton modes $\Delta = \varepsilon^\textrm{ph}_0 - E_g$ (here $\varepsilon^\textrm{ph}_0$ is the cavity photon energy at normal incidence and $E_g$ the exciton bandgap energy) and the Rabi splitting between the upper (UP) and lower (LP) polariton branches.
However, the general approach consists of dividing the LP spectrum into the parabolic photon-like ``condensate'' part in the region $p\sim0$, and the high-energy exciton-like states treated as a ``reservoir'' which is needed to support the existence of this condensate. This approach captures the dynamics of single-energy polariton condensates, governed by the driven-dissipative Gross-Pitaevskii equation (GPE), very well  \cite{wouters2007,berloff,manni,haug14}. At the same time, such description neglects polaritons with varying $p>0$, which feature the momentum-dependent change of the exciton-photon ratio that influences the effective mass, lifetime, and the polariton-polariton interaction. Recently, effort has been taken to derive the modified version of the dissipative GPE renormalized by reservoir-bogolon scattering \cite{hybridBGP}, as well as to include the reservoir-bogolon and bogolon-bogolon scattering in terms of Boltzmann kinetic equations into description of the polariton relaxation \cite{haug2020}.

In the general case, for polariton gases, the excitation spectrum is expected to be different from the equilibrium Bogoliubov dispersion due to decaying nature of the system and presence of the excitonic reservoir \cite{wouters2007,hybridBGP}. One theoretical possibility is that such dissipative spectrum of excitations becomes complex, with the real part being either gapped or diffusive in the region of small momenta \cite{yamamoto2012,ostrovskaya2014}, preventing one from regularly defining the Bogoliubov sound velocity $c_s$. However, the momentum range where this behavior could be observed shrinks with the growth of the polariton lifetime, which leads to the requirement of very large condensates where such a non-sonic behavior could be resolved. On the other hand, including the reservoir-bogolon scattering within the hybrid Bolzmann--Gross--Pitaevskii model \cite{hybridBGP} allows to recover linear, though damped, Bogoliubov modes at low wave vectors, which is supported by the experimental evidence to date \cite{utsunomiya,kohnle,pieczarka2015,stepanov,ballarini2020}.
A more detailed study of the Bogoliubov spectrum branches population mechanisms \cite{pieczarka2020} revealed that the observations deviate from the expected dependencies for negative (photonic) detunings, indicating the influence of non-equilibrium effects, whereas they are consistent with the equilibrium theory for positive (excitonic) detunings.
At the same time there is a discrepancy between the expected and observed sound velocity \cite{kohnle,pieczarka2015,estrecho2021}, not explained by either the textbook Bogoliubov theory \cite{pit_str} or its dissipative modifications, which is ascribed to the significant influence of the reservoir and finite temperature effects. It is a call for further investigations and better understanding of the polariton collective excitations behavior.

The purpose of this paper is therefore to examine the polariton condensate and its excitations at nonzero temperatures, in the Hartree--Fock--Bogoliubov (HFB) approximation, treating the non-condensate particles up to the exciton-like (reservoir) part of the dispersion self-consistently.
Notably, the HFB theory has previously been applied to the coupled photon-exciton system \cite{sarchi} to describe the onset of the polariton off-diagonal long-range order at low densities.
Here, we consider densities well in the ``condensed'' region of the phase diagram of Ref.~\cite{sarchi}, deriving the corrections to the usual Bogoliubov theory due to finite temperatures, full non-parabolic polariton spectrum, and the momentum-dependent interactions of the particles. The sound velocity obtained within our theory is shown to be in good agreement with the existing experimental data.
We address the applicability of the HFB theory at $T\neq0$ in the domain of the exciton interaction strength and the detuning to Rabi splitting ratio, and suggest an intuitive way to stitch the HFB description of Bose condensation to the hydrodynamic description of superfluid behavior, which is applicable in the regime of intermediate correlations ({\it i.e.} at elevated densities) \cite{voronova_PRL,boronat}. This allows to obtain the critical transition temperatures, both for condensation and quasi-condensation, in all ranges of parameters in consideration. We show that at large positive detunings that ensure longer particle lifetimes and allow to describe the system in the assumption of equilibrium \cite{deng2006}, the critical temperature stays high compared to the temperatures of the exciton Bose condensation \cite{butov}. Furthermore, we discuss the integral polariton lifetime dependency on the detuning, and the influence of presence of dark excitons.

\section{Hartree--Fock--Bogoliubov approximation}\label{HFB_theory}

Lower-polariton thermalisation towards equilibrium with the surrounding medium requires faster relaxation towards thermal distributions and slower particle decay, which was experimentally shown to be reached with increasing positive photon-exciton detunings \cite{deng2006}.
The detuning directly controls the exciton fraction in the polariton:
\begin{equation}\label{Xp}
X_p^2 = \frac{1}{2}\left(1 + \frac{\Delta_p}{\sqrt{(\hbar\Omega)^2 + \Delta_p^2}}\right)\!, \end{equation}
where $\Delta_p = \Delta + p^2/2m_\textrm{ph} - p^2/2m_\textrm{ex}$, $m_\textrm{ph} =
\varepsilon^\textrm{ph}_0\epsilon/c^2$ and $m_\textrm{ex}$ are the photon and exciton effective masses, respectively, $c$ is the velocity of light in vacuum, $\epsilon$ the dielectric constant, and $\hbar\Omega$ denotes the Rabi splitting at zero momentum and $\Delta$.
The LP and UP particle operators are given by
$$\hat{P}_{\bf p} = X_p\hat{Q}_{\bf p} + \sqrt{1-X_p^2}\hat{C}_{\bf p},\,\,\hat{U}_{\bf p} = -\sqrt{1-X_p^2}\hat{Q}_{\bf p} + X_p\hat{C}_{\bf p},$$
respectively, with $\hat{Q}_{\bf p}$ and $\hat{C}_{\bf p}$ being the annihilation operators of the exciton and cavity photon modes.
When $X_p^2$ are large, the LP--phonon and LP--LP scattering rates towards low-energy states increase \cite{RMP2010}, allowing lower polaritons to thermalise faster than their lifetime which is governed mainly by photons:
\begin{equation}\label{tau_p}
\tau_\textrm{LP}(p) = \frac{\tau_\textrm{ph}}{1-X_p^2}.
\end{equation}
Keeping that in mind, as well as the drastic increase in microcavities state of the art assuring photon lifetimes $\tau_\textrm{ph}$ of tens to hundreds of picoseconds, we will build an equilibrium finite-temperature theory for a uniform exciton-polariton system with $\Delta\geq0$, basing on the standard bosonic case \cite{pit_str,yukalov}.

\begin{figure}[b]
  \includegraphics[width=0.97\linewidth]{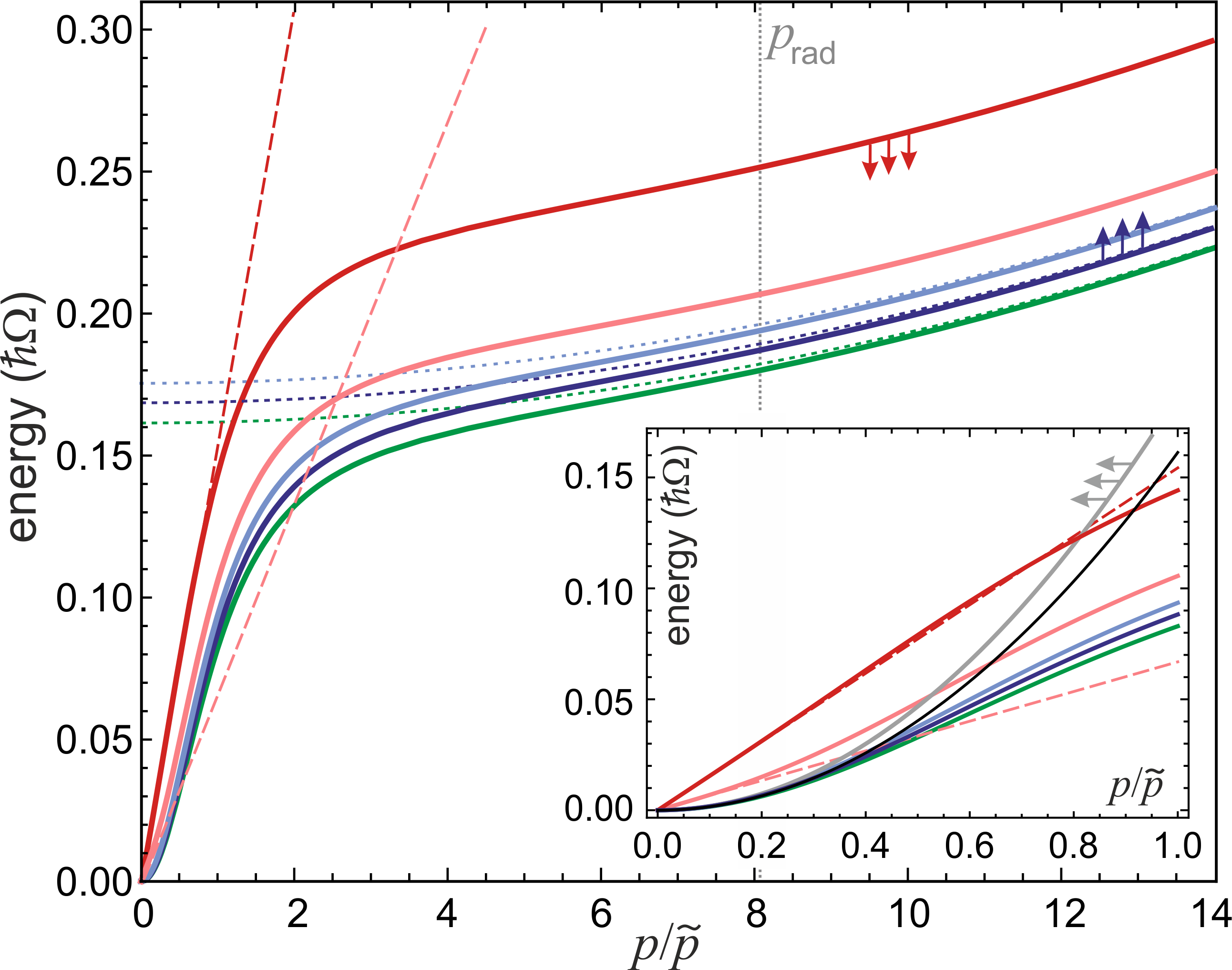}
  \linespread{1.0}\caption{(Color online) 
  Solid lines: the bare LP dispersion $\varepsilon_p$ (green), single-particle spectrum $\varepsilon_p^B$ renormalized by interactions at $T=0$ (dark blue) and at $T=20$~K (light blue), collective excitations spectrum $E_p$ at $T=0$ (dark red) and at $T=20$~K (pink). For single particle dispersions, dotted lines of the corresponding colors show the asymptotic behavior at $p\gg\tilde{p}$, as given in the text. For the Bogoliubov spectra, dashed lines of the corresponding colors indicate the linear law $c_sp$ at $p\to0$. Vertical dotted line shows the light cone boundary $p_\textrm{rad}=E_g\sqrt{\epsilon}/c$. Inset: same, for the enlarged region of low momenta, with additional parabolic asymptotic behaviors $p^2/2m_\textrm{LP}$ (black solid line) and $p^2/2m_\textrm{B}$ (gray solid line) at $p\ll\tilde{p}$. Small arrows indicate the trend for the dispersions change with temperature. All spectra are plotted for the following physical parameters: $E_g=1.6$~eV, $\epsilon=13$, $m_\textrm{ex}=0.22m_0$, $\hbar\Omega=7.2$~meV, $\Delta=10$~meV, $n=6\times10^{10}$~cm$^{-2}$, $g=1.0~\mu$eV~$\mu$m$^2$.}
\label{fig1_spectra}
\end{figure}

We consider the polariton system assuming that both the detuning and the Rabi splitting are small compared to the exciton bandgap energy: $\Delta\ll E_g$, $\hbar\Omega\ll E_g$, and that the UP branch at low temperatures is not populated.
In this case
\begin{equation}\label{QC_p}
\hat{Q}_{\bf p} = X_p\hat{P}_{\bf p}, \quad \hat{C}_{\bf p} = \sqrt{1-X_p^2}\hat{P}_{\bf p}.
\end{equation}
The bare LP spectrum, counted from the bottom of the dispersion, reads
\begin{equation}\label{epsilon_p}
\varepsilon_p = E_0 + \frac{\Delta_p}{2} - \frac{1}{2}\sqrt{\Delta_p^2 + (\hbar\Omega)^2} + \frac{p^2}{2m_\textrm{ex}},
\end{equation}
where $E_0 = [\sqrt{(\hbar\Omega)^2 + \Delta^2}-\Delta]/2$. Expression (\ref{epsilon_p}) allows to derive for $p\to0$ the known effective polariton mass dependence on the detuning \cite{RMP2010}:
\begin{equation}\label{m_LP}
     \frac{1}{m_\textrm{LP}} = \frac{1}{2 m_\textrm{ph}}\left( 1 - \frac{\Delta}{\sqrt{\Delta^2+ (\hbar \Omega)^2}}\right).
\end{equation}
An example of the LP dispersion (\ref{epsilon_p}) is shown in Fig.~\ref{fig1_spectra} by the green solid line. Introducing the characteristic momentum $\tilde{p} = \sqrt{2m_\textrm{LP}E_0}$, one obtains the parabolic dispersions in the two limiting cases $\varepsilon_p \approx p^2/2m_\textrm{LP}$ for $p\ll\tilde{p}$ (black solid line in the inset of Fig.~\ref{fig1_spectra}) and $\varepsilon_p \approx E_0 + p^2/2m_\textrm{ex}$ for $p\gg\tilde{p}$ (green dotted line). Both the polariton mass $m_\textrm{LP}$ and the depth of the ``polariton well'' $E_0$ can be widely tuned. While for negative and zero detunings polaritons stay photon-like, positive $\Delta$ yield two essentially different cases. For $\Delta\lesssim\hbar\Omega$ one has $E_0\gtrsim \hbar\Omega$ and the Hopfield coefficient $X_0^2\sim 1-X_0^2$, providing $m_\textrm{LP} \sim m_\textrm{ph}$. On the contrary, for $\Delta\gg\hbar\Omega$ one gets a shallow polariton well $E_0\approx(\hbar\Omega)^2/4\Delta\ll\hbar\Omega$, and the polaritons become exciton-like $X_0^2\approx1-(\hbar\Omega)^2/4\Delta^2\simeq1$, with the effective mass $m_\textrm{LP} \gg m_\textrm{ph}$.

Since one of our major goals is to account for finite temperatures, the modifications produced in the Bogoliubov description (at $T=0$) by the full non-parabolic polariton dispersion (\ref{epsilon_p}) and momentum-dependent interactions are discussed in Appendix~\ref{AppBog}. Here, we focus on the case $T>0$. Following the procedure introduced by Griffin~\cite{griffin} for atomic gases, we start with the LP Hamiltonian
\begin{multline}\label{hamiltonian_HFB}
\hat{H} = \int\hat{P}^\dag({\bf r})\bigl[\varepsilon(-i\hbar\nabla) - \mu\bigr] \hat{P}({\bf r})d{\bf r} \\
+\frac{1}{2}\int\hat{Q}^\dag({\bf r})\hat{Q}^\dag({\bf r}^\prime)U({\bf r} - {\bf r}^\prime)\hat{Q}({\bf r}^\prime)\hat{Q}({\bf r})d{\bf r}d{\bf r}^\prime \\
 +\int\hat{\tilde{Q}}^\dag({\bf r})\hat{Q}^\dag({\bf r}^\prime) \tilde{U}({\bf r}-{\bf r}^\prime)\hat{Q}({\bf r}^\prime)\hat{\tilde{Q}}({\bf r})d{\bf r}d{\bf r}^\prime,
\end{multline}
where $\varepsilon(-i\hbar\nabla)$ is given by (\ref{epsilon_p}) with the substitution of the first-quantized momentum operator, $\mu$ is the chemical potential of the system,
\begin{equation}\label{PQ}
\hat{P}({\bf r}) = \frac{1}{\sqrt{S}}\sum\limits_{\bf p}e^{\frac{i}{\hbar}{\bf p}\cdot{\bf r}} \hat{P}_{\bf p},\quad\hat{Q}({\bf r}) = \int X({\bf r}-{\bf r}^\prime)\hat{P}({\bf r}^\prime)d{\bf r}^\prime
\end{equation}
are the lower polariton and the exciton field operators, respectively ($S$ being the quantization area). Here we have introduced the notation
\begin{equation}
X({\bf r}-{\bf r}^\prime\!) = \frac{1}{S}\sum\limits_{\bf p}e^{\frac{i}{\hbar}{\bf p}\cdot({\bf r}-{\bf r}^\prime\!)} X_p, \,\, \int X({\bf r})d{\bf r} = X_0,
\end{equation}
where one can switch to integration $\frac{1}{S}\sum\limits_{\bf p}\rightarrow\int \frac{d{\bf p}}{(2\pi\hbar)^2}$, since we define $X_p$ as the positive square root of the r.h.s. of (\ref{Xp}).
In (\ref{hamiltonian_HFB}), we have additionally taken into account the interaction of exciton-polaritons with dark excitons, whose field operator is denoted as $\hat{\tilde{Q}}({\bf r})$. In the general case, the interaction between the bright and dark excitons $\tilde{U}(r)$ does not coincide with the bright exciton-exciton interaction $U(r)$, since the scattering length $a_s$ is spin-dependent. We assume that the dark excitons do not convert into bright excitons $[\hat{\tilde{Q}}({\bf r}),\hat{Q}^\dag({\bf r})]=0$, and hence do not take part in the condensation process, while they still contribute to the chemical potential \cite{darkex}.

Using the Hamiltonian (\ref{hamiltonian_HFB}) and the commutation relation $[\hat{P}({\bf r}),\hat{Q}^\dag({\bf r})]=X({\bf r}-{\bf r}^\prime)$, one gets the Heisenberg equation for the polariton field operator:
\begin{multline}\label{heisenberg}
i\hbar\frac{\partial}{\partial t}\hat{P}({\bf r},t) = \bigl[\varepsilon(-i\hbar\nabla) - \mu\bigr] \hat{P}({\bf r},t) \\
+ \int\!\! X({\bf r}\!-{\bf r}^\prime)\hat{Q}^\dag({\bf r}^{\prime\prime}\!\!,t)U({\bf r}^\prime - {\bf r}^{\prime\prime})\hat{Q}({\bf r}^{\prime\prime}\!\!,t)\hat{Q}({\bf r}^\prime\!,t)d{\bf r}^\prime d{\bf r}^{\prime\prime} \\
+ \int\! X({\bf r}-{\bf r}^\prime)\hat{\tilde{Q}}^\dag({\bf r}^{\prime\prime}\!\!,t)\tilde{U}({\bf r}^\prime\! - {\bf r}^{\prime\prime}) \hat{\tilde{Q}}({\bf r}^{\prime\prime}\!\!,t)\hat{Q}({\bf r}^\prime\!,t)d{\bf r}^\prime d{\bf r}^{\prime\prime}\!\!.
\end{multline}

To rewrite Eq.~(\ref{heisenberg}) in the Hartree--Fock approach, we separate the condensate in both fields \cite{griffin,beliaev},
\begin{equation} \label{bog_subs}
    \hat{P}({\bf r},t) = \sqrt{n_0} + \hat{P}^\prime({\bf r},t), \quad
    \hat{Q}({\bf r},t) = X_0 \sqrt{n_0} + \hat{Q}^\prime({\bf r},t),
\end{equation}
where $n_0$ is the LP condensate density,
\begin{equation}\label{Q-prime}
  \hat{Q}^\prime({\bf r},t) = \int\! X({\bf r}-{\bf r}^\prime)\hat{P}^\prime({\bf r}^\prime, t)d{\bf r}^\prime,
\end{equation}
and the average with the Gibbs density matrix $\langle \hat{P}^\prime({\bf r},t)\rangle = \langle \hat{Q}^\prime({\bf r},t)\rangle = 0$. Using (\ref{bog_subs}) and the self-consistent HFB approximation
\begin{multline}\label{HFB}
  \hat{Q}^{\prime\dag}({\bf r}^{\prime\prime})\hat{Q}^\prime({\bf r}^{\prime\prime})
\hat{Q}^\prime({\bf r}^\prime) = \langle\hat{Q}^{\prime\dag}({\bf r}^{\prime\prime})\hat{Q}^\prime({\bf r}^{\prime\prime})\rangle\hat{Q}^\prime({\bf r}^\prime) \\ +\hat{Q}^\prime({\bf r}^{\prime\prime})\langle \hat{Q}^{\prime\dag}({\bf r}^{\prime\prime})\hat{Q}^\prime({\bf r}^\prime)\rangle + \hat{Q}^{\prime\dag}({\bf r}^{\prime\prime})\langle\hat{Q}^\prime({\bf r}^{\prime\prime})
\hat{Q}^\prime({\bf r}^\prime)\rangle
\end{multline}
(the time variable is omitted for clarity) brings the second term in (\ref{heisenberg}) to the mean-field form
\begin{multline}\label{3oper_int}
  \hat{Q}^\dag({\bf r}^{\prime\prime})\hat{Q}({\bf r}^{\prime\prime})\hat{Q}({\bf r}^\prime) = \\
  = X_0\sqrt{n_0}\left[X_0^2n_0 + n^\prime_Q + \rho^\prime_{1Q}({\bf r}^{\prime\prime},{\bf r}^\prime) + m^\prime_Q({\bf r}^\prime,{\bf r}^{\prime\prime})\right] \\
  +\hat{Q}^\prime({\bf r}^\prime)\left(X_0^2n_0 + n^\prime_Q\right) +
  \hat{Q}^\prime({\bf r}^{\prime\prime})\left(X_0^2n_0 + \rho^\prime_{1Q}({\bf r}^{\prime\prime},{\bf r}^\prime)\right) \\
  +\hat{Q}^{\prime\dag}({\bf r}^{\prime\prime})\left(X_0^2n_0 + m_Q^\prime({\bf r}^{\prime\prime},{\bf r}^\prime)\right).\qquad\qquad\qquad
\end{multline}
The third term in (\ref{heisenberg}) in the same approximation is
\begin{equation}\label{dark_HFB}
  \hat{\tilde Q}^\dag({\bf r}^{\prime\prime})\hat{\tilde Q}({\bf r}^{\prime\prime})\hat{Q}({\bf r}^\prime) = \left[X_0\sqrt{n_0} + \hat{Q}^\prime({\bf r}^\prime)\right]\tilde{n}.
\end{equation}
In (\ref{3oper_int}), (\ref{dark_HFB}), the following notations have been introduced for non-condensate exciton density and one-body density matrix, the anomalous average, and the density of dark excitons, respectively:
\begin{eqnarray}
  n^\prime_Q &\equiv& \langle\hat{Q}^{\prime\dag}({\bf r})\hat{Q}^\prime({\bf r})\rangle =  \int\!\! X_p^2\langle\hat{P}_{\bf p}^\dag\hat{P}_{\bf p}\rangle \frac{d{\bf p}}{(2\pi\hbar)^2},\label{n'} \\
  \rho_{1Q}^\prime({\bf r},{\bf r}^\prime)\! &\equiv& \langle\hat{Q}^{\prime\dag}({\bf r})\hat{Q}^\prime({\bf r}^\prime)\rangle = \!\!\int\!\! X_p^2 e^{\frac{i}{\hbar}{\bf p}\cdot({\bf r}^\prime\!-{\bf r})} \langle\hat{P}_{\bf p}^\dag\hat{P}_{\bf p}\rangle \frac{d{\bf p}}{(2\pi\hbar)^2}, \nonumber \\
  m^\prime_Q({\bf r},{\bf r}^\prime)\! &\equiv& \langle\hat{Q}^\prime({\bf r})\hat{Q}^\prime({\bf r}^\prime)\rangle = \!\!\int\!\! X_p^2 e^{\frac{i}{\hbar}{\bf p}\cdot({\bf r}^\prime\!-{\bf r})} \langle\hat{P}_{\bf p}\hat{P}_{\bf -p}\rangle \frac{d{\bf p}}{(2\pi\hbar)^2}, \nonumber \\
  \tilde{n} &\equiv& \langle\hat{\tilde Q}^\dag({\bf r})\hat{\tilde Q}({\bf r})\rangle.\label{n''}
\end{eqnarray}
In fact, since we have neglected the spin flip processes between the dark and bright excitons, the theory presented below is applicable also for the case when the density $\tilde{n}$ is that of any background particles (such as electrons, trions, incoherent excitons, {\it etc.}) that principally do not convert into polaritons, but still interact with their excitonic fraction via the potential $\tilde{U}({\bf r})$. 
It is important to note that the background particles (electrons in particular) should be of low density compared to the density of polaritons, so as not to influence the exciton-exciton interaction. For the case of doped semiconductors where the electron densities can be made high, polariton-electron interaction gets enhanced by light-matter coupling, as shown recently in \cite{meera} for 2D transition-metal dichalcogenides (TMDs). We consider the regime where such renormalizations are negligible.

To include $\tilde{n}$ in our calculations quantitatively, within the equilibrium description we assume that this background population is that of excitons of $s-1$ spin degrees of freedom, {\it i.e.}
\begin{equation}\label{darkexcitons}
\tilde{n} = \sum\limits_{\sigma=2}^s \int \frac{1}{e^{(E_0 + p^2/2m_\textrm{ex})/T}-1}\frac{d{\bf p}}{(2\pi\hbar)^2}.
\end{equation}
In particular, for GaAs $s=4$, with $\sigma=1$ corresponding to the Bose-condensed polaritons, $\sigma=2$ to the second bright branch that does not undergo condensation, and $\sigma=3$,~4 to dark excitons. Estimates show that the branch $\sigma=2$, despite being coupled to light, is occupied mostly in the high-momenta excitonic region $p>\tilde{p}$, so that their dispersion law in (\ref{darkexcitons}) is taken coinciding with the exciton dispersion \cite{brightex}.
The current assumptions \cite{darkex} work well in the regime of continuous-wave pumping,
allowing to treat total density $n$ of polaritons as experimentally controllable quantity even when there are equilibrium species populating the other branches $\sigma=2,3,\dots,s$.

Following Popov \cite{popov}, we discard the anomalous non-condensate term: $m_Q^\prime({\bf r},{\bf r}^\prime)=0$. However, in contrast to Ref.~\cite{popov}, it is justified not by the smallness of the condensate fraction $n_0/n$ which is large in our case, but by fact that the exciton pair interaction becomes dressed up to the Beliaev ladder (a more detailed discussion is provided in Appendix~\ref{AppM}).
Within the introduced framework, averaging the Heisenberg equation (\ref{heisenberg}) yields the expression for the chemical potential of polaritons:
\begin{multline}\label{mu}
    \!\! \mu \!=\! X_0 \!\!\int\!\! X({\bf r}^\prime)U({\bf r}^\prime - {\bf r}^{\prime\prime}\!)\!\left[X_0^2 n_0 + n_Q^\prime \!+ \rho^\prime_{1Q}({\bf r}^{\prime\prime}\!\!, {\bf r}^\prime)\!\right]\!d{\bf r}^\prime d{\bf r}^{\prime\prime} \\
    + X_0\tilde{n}\!\!\int\!\! X({\bf r}^\prime)\tilde{U}({\bf r}^\prime - {\bf r}^{\prime\prime})d{\bf r}^\prime d{\bf r}^{\prime\prime}.
\end{multline}
In the case of contact interaction for both bright and dark excitons, $U({\bf r})=g\delta({\bf r})$, $\tilde{U}({\bf r})=\tilde{g}\delta({\bf r})$, the expression (\ref{mu}) reduces to
\begin{equation}\label{mu_contact}
\mu = gX_0^2(X_0^2n_0 + 2n_Q^\prime) + \tilde{g}X_0^2\tilde{n}.
\end{equation}
For the sake of generality, however, we assume for all derivations, unless stated otherwise, the exciton-exciton pair interaction to 
have a general (not delta-functional) shape. This makes all the formulae applicable to the cases when interaction cannot be approximated as contact, {\it e.g.}, when large momenta are considered or when one deals with dipolar excitons. In the following we will denote $g = U(0) \neq U({\bf p}\neq0)$, where $U({\bf p})$ is the Fourier image of $U({\bf r})$.

\begin{figure}[t]
  \includegraphics[width=1.00\linewidth]{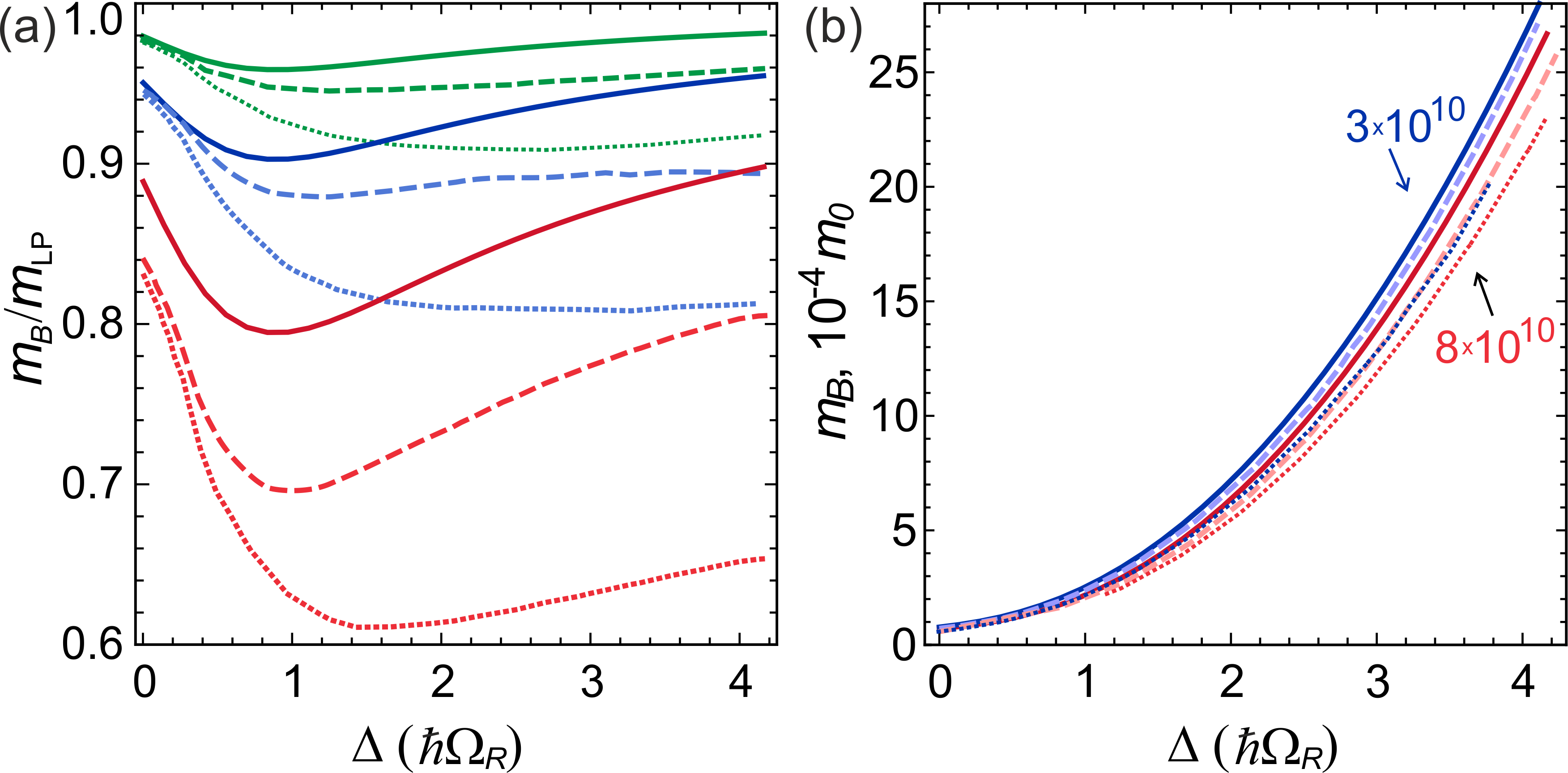}
  \linespread{1.0}\caption{(Color online)
  {\bf a}, Change of the renormalized particle effective mass relative to the lower polariton mass $m_B/m_\textrm{LP}$ for $T=0$ (solid lines) and $T=10$~K (dashed lines), depending on the energy detuning, for three values of the interaction strength $g=1$ (green), $2.5$ (blue), and $6~\mu$eV~$\mu$m$^2$ (red lines) and the total density $n=3\times10^{10}$~cm$^{-2}$.
  {\bf b}, Absolute value of $m_B$ depending on the detuning, for $T=0$ (solid) and $T=10$~K (dashed lines), for two values of total polariton density as marked, for $g=2.5~\mu$eV~$\mu$m$^2$.
  For both {\bf a} and {\bf b}, the dotted lines of the corresponding colors show the same dependencies when the density $\tilde{n}$ of dark excitons is included in consideration, with $\tilde{g}=g$. Other parameters as in Fig.~\ref{fig1_spectra}.}
\label{fig2_m_B}
\end{figure}

Subtracting from Eq.~(\ref{heisenberg}) its averaged version, one gets the equation for the non-condensed fraction of the polariton field $\hat{P}^\prime({\bf r}, t)$:
\begin{multline}\label{P-prime-eq}
i \hbar \frac{\partial}{\partial t}\hat{P}^\prime({\bf r}, t)= \left[\varepsilon (-i\hbar\nabla) - \mu\right]\hat{P}^\prime({\bf r}, t) \\
 + \int X({\bf r} - {\bf r}^\prime) U({\bf r}^\prime - {\bf r}^{\prime\prime}) \left[ \hat{Q}^{\prime}({\bf r}^\prime,t)(X_0^2 n_0 + n^\prime_Q)\right. \\
 \left. + \hat{Q}^\prime({\bf r}^{\prime\prime}, t)(X_0^2 n_0 + \rho^\prime_{1Q}({\bf r}^{\prime\prime}, {\bf r}^\prime)) + \hat{Q}^{\prime\dag}({\bf r}^{\prime\prime}, t)X_0^2n_0 \right] d{\bf r}^\prime d{\bf r}^{\prime\prime} \\
 + \tilde{g}\tilde{n}\int X({\bf r} - {\bf r}^\prime)\hat{Q}^{\prime}({\bf r}^\prime,t)d{\bf r}^\prime,\qquad\qquad
\end{multline}
where $\hat{Q}^\prime({\bf r},t)$ is given by (\ref{Q-prime}). After the Fourier transform
$$\hat{P}^\prime({\bf r}) = \frac{1}{\sqrt{S}} \sum\limits_{{\bf p}\neq0} \hat{P}_{\bf p} e^{\frac{i}{\hbar}{\bf p\cdot r}}, \,\,
\hat{Q}^\prime({\bf r}) = \frac{1}{\sqrt{S}} \sum\limits_{{\bf p}\neq0}  X_p\hat{P}_{\bf p} e^{\frac{i}{\hbar}{\bf p\cdot r}},$$
Eq.~(\ref{P-prime-eq}) takes a simple form:
\begin{equation}\label{Pp}
  i\hbar\frac{\partial}{\partial t} \hat{P}_{\bf p}(t) = \bigl(\varepsilon_p^B + \mu_p\bigr) \hat{P}_{\bf p}(t) + \mu_p\hat{P}^\dag_{-\bf p}(t),
\end{equation}
with the renormalized single-particle spectrum
\begin{multline}\label{epsilon_B}
  \varepsilon_p^B = \varepsilon_p + \left(X_p^2-X_0^2\right)\left[g(X_0^2n_0+ n^\prime_Q) + \tilde{g}\tilde{n}\right] \\
  + \!\!\int\!\! X_{p^\prime}^2 \bigl[X_p^2 U({\bf p}-{\bf p}^\prime) -X_0^2 U({\bf p}^\prime)\bigr] \langle\hat{P}^{\dag}_{\bf p^\prime} \hat{P}_{\bf p^\prime}\rangle \frac{d{\bf p}^\prime} {(2\pi\hbar)^2}
\end{multline}
and
\begin{equation}\label{mu_p}
  \mu_p = U({\bf p})X_0^2X_p^2n_0.
\end{equation}
The superscript `$B$' in (\ref{epsilon_B}) and below stands for `Bogoliubov', denoting that the renormalizations are derived within the HFB approximation (\ref{HFB}).
The appearance of the spectrum $\varepsilon_p^B$ in the Heisenberg equation (\ref{Pp}) for the LP field operator instead of the bare spectrum $\varepsilon_p$ is due to the fact that polariton interaction contains an extra dependence on momentum compared to the exciton interaction: $U_\textrm{LP}({\bf p}) = X_0^2X_p^2U({\bf p})$.
Correspondingly, both the depth of the polariton well and the effective mass get renormalized:
\begin{equation}\label{epsilon_B_asymp}
\varepsilon_p^B \approx
\left\{
\begin{array}{lc}
  p^2/2m_B, & p\ll\tilde{p} \\
  E_0 + \mu(1-X_0^2)/X_0^2 + p^2/2m_\textrm{ex}, & p\gg\tilde{p}
\end{array}
\right.
\end{equation}
\begin{equation}\label{m_B}
\frac{1}{m_B} = \frac{1}{m_\textrm{LP}}\left(1 + \frac{2\mu}{\sqrt{\Delta^2 + (\hbar\Omega)^2}}\right).
\end{equation}
This renormalization is present both at zero and finite $T$. It should be noted that, while in the second line of (\ref{epsilon_B_asymp}) the chemical potential $\mu$ is given by (\ref{mu}), in (\ref{m_B}) it should be taken in the shape (\ref{mu_contact}), as $U({\bf p})\approx U(0)$ for $p\ll\tilde{p}$. The example of the spectrum $\varepsilon_p^B$ is plotted in Fig.~\ref{fig1_spectra} by the dark blue (for $T=0$) and light blue (for $T=20$~K) solid lines. One can see that, while for $p\rightarrow0$ the dispersions $\varepsilon_p$, $\varepsilon_p^{B(T=0)}$ (given by (\ref{eps_bog_T0})) and $\varepsilon_p^B$ differ negligibly, for higher momenta $p\gtrsim\tilde{p}$, the difference starts to play a role. The ratio of the renormalized particle effective mass $m_B$ to $m_\textrm{LP}$ is less than unity and it is non-monotonous with respect to $\Delta$, displaying a pronounced minimum for larger interactions, as shown in Fig.~\ref{fig2_m_B}a.
However, since $m_\textrm{LP}$ is itself detuning-dependent (see (\ref{m_LP})), the dependence of $m_B$ on $\Delta$ is regular, with lower slopes for higher temperatures and larger densities (see Fig.~\ref{fig2_m_B}b).

From (\ref{Pp}), using the Bogoliubov transformation
\begin{equation}\label{bog_trans}
\hat{P}_{\bf p}(t) = u_p\hat{\alpha}_{\bf p}(t)-v_p\hat{\alpha}_{\bf -p}^\dag(t),\,\, \hat{P}_{\bf p}^\dag(t) = u_p\hat{\alpha}_{\bf p}^\dag(t) - v_p\hat{\alpha}_{\bf -p}(t)
\end{equation}
with $u_p^2-v_p^2 = 1$ and
$$\hat{\alpha}_{\bf p}(t) = \hat{\alpha}_{\bf p} e^{-\frac{i}{\hbar}E_pt},\quad \hat{\alpha}_{\bf -p}^\dag(t) = \hat{\alpha}_{\bf -p}^\dag e^{\frac{i}{\hbar}E_pt},$$
one gets the spectrum of elementary excitations of the polariton Bose gas in HFB approximation:
\begin{equation}\label{bog_spectrum}
E_p = \sqrt{\varepsilon_p^B(\varepsilon_p^B + 2\mu_p)}
\end{equation}
and the Bogoliubov coefficients
\begin{equation}\label{uv_bog}
u_p^2,v_p^2 \!=\! \frac{(E_p \!\pm\! \varepsilon_p^B)^2}{4\varepsilon_p^BE_p} \!=\! \frac{1}{2}\!\left(\!\!\sqrt{1\!+\!\frac{\mu_p^2}{E_p^2}}\pm1\!\!\right) \!=\! \frac{\varepsilon_p^B\!+\!\mu_p\!\pm\! E_p}{2E_p}.
\end{equation}

The main general results of our consideration so far are the analytical expressions for chemical potential (\ref{mu}), renormalized particle spectrum (\ref{epsilon_B}), and the spectrum of elementary excitations (\ref{bog_spectrum}). In all expressions the influence of finite temperature is implicitly included via the non-condensate exciton density $n^\prime_Q$ and the background dark density $\tilde{n}$. Calculating the occupation number
\begin{equation}\label{n_p}
  n_p = \langle\hat{P}^{\dag}_{\bf p} \hat{P}_{\bf p}\rangle = v_p^2 + \frac{2v_p^2}{e^{E_p/T}-1} + \frac{1}{e^{E_p/T}-1},
\end{equation}
where the temperature $T$ is expressed in energy units, one gets the following set of integral equations (see (\ref{n'})):
\begin{equation}\label{int_system}
   \left\{
    \begin{array}{lc}
    \displaystyle n_0 + \int n_p \frac{d{\bf p}}{(2\pi\hbar)^2} = n,\\[6pt]
    \displaystyle n^\prime_Q - \int X_p^2 n_p \frac{d{\bf p}}{(2\pi\hbar)^2} = 0.
    \end{array}
   \right.
\end{equation}
Solving the Eqs.~(\ref{int_system}) together with (\ref{n_p}) allows to find the polariton condensate and non-condensate densities $n_0$ and $n^\prime = \int n_p d{\bf p}/(2\pi\hbar)^2$, and obtain $\varepsilon_p^B$, $m_B$, and $E_p$ according to (\ref{bog_spectrum}) quantitatively, for each value of the detuning $\Delta$ and injected polariton density $n$ at different temperatures. The exemplary results of such calculations are shown in the above Figures~\ref{fig1_spectra} and~\ref{fig2_m_B} for $\Delta=10$~meV (at the Rabi splitting $\hbar\Omega = 7.2$~meV). In particular, one can see in Fig.~\ref{fig1_spectra} that the spectrum of collective excitations, shown by red solid lines, at finite $T$ shifts considerably not only in the region of large momenta (like $\varepsilon_p^B$ compared to $\varepsilon_p$), but also at $p\ll\tilde{p}$. Change of the slope of the linear part of the dispersion (red dashed lines) results in the change of the Bogoliubov sound velocity (see below in more details). Finally, solving the Eqs.~(\ref{int_system}) at different $T$ and $n$ allows to obtain the critical temperature of Bose condensation $T_C$ (defined as the temperature at which $n_0\to0$) for a finite polariton system, and its dependence on the detuning.

\section{Results and discussion}

Prior to addressing the critical temperature of transition to the Bose-condensed state and the applicability of the developed description, we focus on the influence of finite temperature on the collective excitations spectrum and its dependence on the polariton density and detuning. For calculations presented below we have used the exciton interaction in the shape of the Lennard--Jones potential $U(r) = \hbar^2/(m_\textrm{ex}x_0^2)[(x_0/r)^{12} - (x_0/r)^6]$ with $x_0=14.14$~nm for the case $T=0$. For finite $T$, for simplicity of calculations using the implicit scheme involving Eqs. (\ref{n_p}), (\ref{int_system}), we used the contact potential $U(r)=g\delta(r)$ with different interaction constants $g$ (from 1 to 6~$\mu$eV~$\mu$m$^2$ \cite{estrecho_TF}). It is important to note however that even within the simplified description the choice of the interaction constant $g$ for each value of $n$ is an open question, especially for the systems allowing high densities, such as TMDs \cite{bleu} or organic polariton condensates \cite{blueshifts}.
Furthermore, for gases of dipolaritons \cite{byrnes_dip} where excitons interact as dipoles, even in the case of weak interactions and low densities the full treatment with the inclusion of $U({\bf p})$ would be essential, since the dipole-dipole interaction is not short-ranged.

\subsection{The Bogoliubov spectrum and sound velocity}

\begin{figure}[b]
  \includegraphics[width=1.00\linewidth]{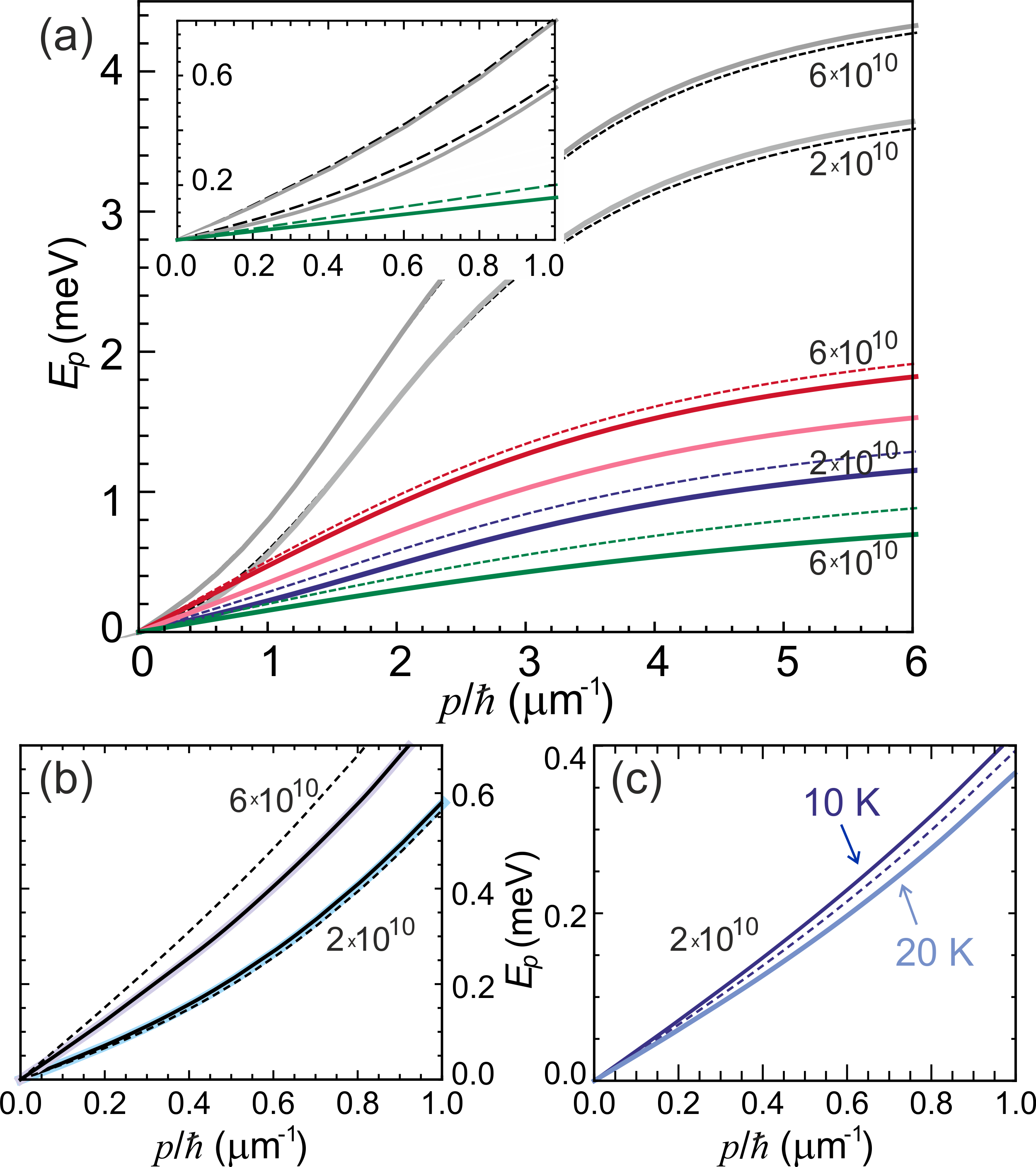}
  \linespread{1.0}\caption{(Color online) 
  {\bf a}, The spectrum of collective excitations (\ref{bog_spectrum}) at $\hbar\Omega = 7.2$~meV for $T=0$ (dashed), $10$~K (dark solid), and $20$~K (light solid lines, where applicable), for energy detunings $\Delta = 0$ (black), $10$~meV (red, blue) and $30$~meV (green), for the total polariton densities $n=2\times10^{10}$~cm$^{-2}$ and $n=6\times10^{10}$~cm$^{-2}$ as marked; $g=2.5~\mu$eV~$\mu$m$^2$. Inset: close-up on the region $p\sim0$, only the lines for $\Delta=0$ and $30$~meV are shown for clarity.
  {\bf b, c}, Low-momenta region of the spectrum $E_p$ for $\hbar\Omega=15.8$~meV, for $\Delta=0$ (b) and $\Delta=10$~meV (c) for total densities as marked. The dashed lines represent the dispersions for $T=0$. In {\bf b}, thin black lines for $T=10$~K and coinciding colored thick lines for $T=20$~K. In {\bf c}, solid lines for the temperatures as marked.  }
\label{fig3_Ep}
\end{figure}

Calculating the asymptotic value of the momentum-energy dispersion (\ref{bog_spectrum}) at large momenta $p\gg\tilde{p}$ for $T>0$ (using (\ref{epsilon_B_asymp})) and $T=0$ (using (\ref{eps_bog_T0})), one sees that the high-momentum tail of $E_p$ can get shifted either up or down with the temperature increase, depending on the detuning, Rabi splitting, and polariton density. In particular, in the case $n^\prime/(2n^\prime_Q+\tilde{n}) < (1-X_0^2)/X_0^2(2-X_0^2)$ which happens at near-zero detunings or at large contributions of $\tilde{n}$, the tail of the dispersion goes up with $T$. This is shown in Fig.~\ref{fig3_Ep}a (for $\hbar\Omega=7.2$~meV) at $\Delta=0$: for both presented values of $n$, the dispersions at $T=20$~K (solid gray lines) go higher at large $p$ than those for $T=0$ (dashed black lines). However with the increase of $\Delta$ this inequality does not hold any longer and all dispersions shift down for higher $T$ (see Fig.~\ref{fig1_spectra} and colored lines in Fig.~\ref{fig3_Ep}a). Critical $\Delta$ at which this change happens depends on $\hbar\Omega$ and $n$, as those implicitly enter the inequality above.

Analysis of the low-momenta region of $E_p$ is not as straightforward. Looking at the asymptotics of the spectrum (\ref{bog_spectrum}) at $p\ll\tilde{p}$, one gets the regular Bogoliubov linearization $E_p(p\rightarrow0)=c_sp$ with
\begin{equation}\label{c_s_T}
     c_s = \sqrt{\frac{\mu -  X_0^2(2gn^\prime_Q + \tilde{g}\tilde{n})}{m_B}}.
\end{equation}
It is important to note that the numerator in (\ref{c_s_T}) differs from the chemical potential $\mu$ (similarly to the result of Ref.~\cite{hybridBGP}), and the denominator contains the renormalized effective mass $m_B$ given by (\ref{m_B}) instead of $m_\textrm{LP}$, due to the presence of finite-temperature contributions $n^\prime_Q$ and $\tilde{n}$. Thus we analytically recover the deviation of the sound velocity from the standard Bogoliubov definition $c_s^\textrm{Bog}=\sqrt{\mu/m_\textrm{LP}}$, which is regularly observed in experiments and is usually attributed to dissipative nature of polaritons \cite{kohnle,pieczarka2015,estrecho2021}.
Indeed, even with the simplifying assumptions of very low temperature $T\to0$ that would result in $\mu\approx gn_0X_0^4$, one gets from (\ref{c_s_T}) the ratio $c_s/c_s^\textrm{Bog} \approx \sqrt{1+2\mu/\sqrt{(\hbar\Omega)^2+\Delta^2}}$. Comparing it to existing experiments~\cite{kohnle,pieczarka2015} which both report $c_s$ extracted from the slope of the dispersion to be higher than $c_s^\textrm{Bog}$ calculated from the measured blueshift, with the parameters given in Ref.~\cite{kohnle} one gets $c_s = 1.138c_s^\textrm{Bog}$, and with the parameters of Ref.~\cite{pieczarka2015} $c_s = 1.265c_s^\textrm{Bog}$.
While the number for $c_s^\textrm{Bog}$ is not given in Ref.~\cite{kohnle}, our result is in close agreement with the numbers reported in Ref.~\cite{pieczarka2015} ($c_s = 1.95~\mu$m/ps vs. $c_s^\textrm{Bog} = 1.45~\mu$m/ps). The increase of the sound velocity compared to the value at $T=0$ is seen in Fig.~\ref{fig3_Ep}b ($\Delta=0$) and in Fig.~\ref{fig3_Ep}c ($\Delta\approx0.6\hbar\Omega$) for $T=10$~K (dark blue line), both for the total density $n=2\times10^{10}$~cm$^{-2}$.
Considering higher temperatures (see Fig.~\ref{fig3_Ep}c, light blue line), higher $\Delta/\hbar\Omega$ (Fig.~\ref{fig3_Ep}a), or higher densities (Fig.~\ref{fig3_Ep}b, $n=6\times10^{10}$~cm$^{-2}$) results in an inverse effect of lowering $c_s$ with respect to $c_s^\textrm{Bog}$, which has also been observed in experiment \cite{estrecho2021}.

\begin{figure}[b]
  \includegraphics[width=1.00\linewidth]{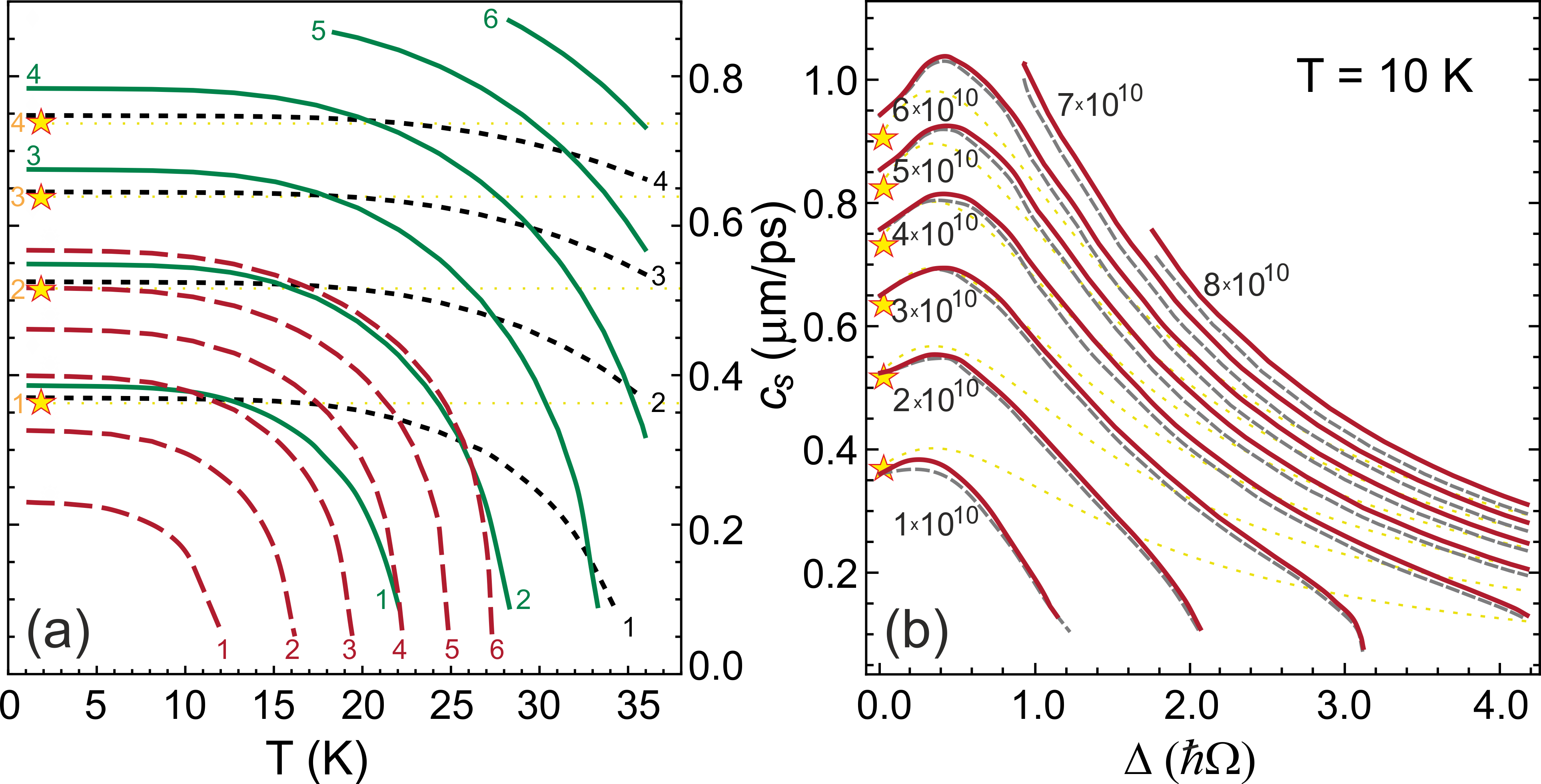}
  \linespread{1.0}\caption{(Color online)
  {\bf a}, The sound velocity $c_s$ according to (\ref{c_s_T}) versus temperature $T$ for $\Delta=0$ (black dotted lines), $10$~meV~$\approx0.6\hbar\Omega$ (green solid lines), and $30$~meV~$\approx1.9\hbar\Omega$ (red dashed lines), for values of the total polariton density $n$ from $1$ to $6\times10^{10}$~cm$^{-2}$ as marked. For this panel, $\hbar\Omega=15.8$~meV, $\tilde{n}=0$.
  {\bf b}, $c_s$ dependent on the detuning $\Delta$ at $T=10$~K, for different polariton densities (as marked), without (dashed) and  with the dark excitons $\tilde{n}$ taken into account (red solid lines), assuming $\tilde{g}=g$. 
  Here, $\hbar\Omega = 7.2$~meV. For both panels, $g=2.5~\mu$eV~$\mu$m$^2$. The yellow stars indicate the value of the standard Bogoliubov sound velocity $c_s^\textrm{Bog}=\sqrt{gnX_0^4/m_\textrm{LP}}$ for the corresponding densities, calculated at $\Delta=0$. In {\bf b}, the yellow dotted lines show $c_s^\textrm{Bog}(\Delta)$ for $n=1$--$6\times10^{10}$~cm$^{-2}$.
  }
\label{fig4_cs}
\end{figure}

To analyse better this change of the slope and hence the sound velocity, in Fig.~\ref{fig4_cs} we plot $c_s$ against $T$ and $\Delta$ for various values of the density $n$. As clearly seen in Fig.~\ref{fig4_cs}a, at low temperatures and not too high detunings, velocity given by (\ref{c_s_T}) is higher than $c_s^\textrm{Bog}$ calculated for the same values of $gn$ (indicated by the yellow marks), whereas at the increase of $T$ the sound velocity is lowered until it drops to zero when the temperature approaches $T_C$ for given $n$ and $\Delta$ (see below). At higher detunings (the red dashed lines for $\Delta\sim2\hbar\Omega$), $c_s$ is considerably lower than $c_s^\textrm{Bog}$ for all values of $n$. Interestingly, there is a peak observed in the dependence of $c_s$ on $\Delta$ in the region of small detunings: all lines for $\Delta=10$~meV in Fig.~\ref{fig4_cs}a at low temperatures go higher than those for $\Delta=0$. The same peak is clearly seen in Fig.~\ref{fig4_cs}b, where $c_s$ is plotted against $\Delta$ at $T=10$~K. Additionally, Fig.~\ref{fig4_cs}b shows the change of the sound velocity when the density of dark excitons is accounted for according to (\ref{darkexcitons}). One sees that for a given set of $(n,\Delta,T)$, the sound velocity increases when the dark population is considered. However, since $\tilde{n}$ also contributes to the chemical potential, with respect to measured blueshift $c_s$ appears effectively decreased. The complication of treating this case analytically is due to the dependence of $n^\prime_Q$, $\tilde{n}$ on both $T$ and the injected density of polaritons $n$, which makes the functional dependence of $c_s$ on $\mu$ rather sophisticated. Such analysis, as well as the consideration of out-of-equilibrium background particles, lies out of the scope of the current work.

\subsection{Integral polariton lifetime} \label{sec_lifetime}

Calculation of the occupation number (\ref{n_p}) allows to find the mean integral lifetime of polaritons in the system at the temperature $T$:
\begin{equation}\label{tau_int}
\frac{1}{\tau} =  \frac{1}{n+\tilde{n}_2}\left[\frac{n_0}{\tau_\textrm{LP}(0)} + \int \frac{ \langle P^\dag_{\bf p} P_{\bf p}\rangle}{\tau_\textrm{LP}(p)}\frac{d{\bf p}}{(2\pi\hbar)^2} + \frac{\tilde{n}_2}{\tau_\textrm{ex}}\right],
\end{equation}
with $\tau_\textrm{LP}(p)$ given by (\ref{tau_p}), $\tilde{n}_2$ being the occupation of the second exciton branch (see (\ref{darkexcitons})) which is also coupled to light \cite{brightex}, and the integration performed up to the edge of the exciton radiative zone, {\it i.e.} up to $p_\textrm{rad}=E_g\sqrt{\epsilon}/c$. In (\ref{tau_p}), we take $\tau_\textrm{ph}=10$~ps to be the same for all momenta, while this may not always hold as the quality factor of microcavities drops with the increase of the photons angle-of-incidence with respect to the cavity normal. At the same time, with the increase of $p$ at large positive detunings, $\tau_\textrm{LP}(p)$ calculated using (\ref{tau_p}) quickly reaches the exciton lifetime $\tau_\textrm{ex}$ (ranging from 0.5 to 1~ns, see, {\it e.g.}, \cite{deng_PNAS}), so that the integrand in the second term of (\ref{tau_int}) should be cut off at the momentum corresponding to $\tau_\textrm{LP}(p^*)=\tau_\textrm{ex}$. From $p^*$ to $p_\textrm{rad}$, $\tau_\textrm{LP}$ is taken constant and equal to $\tau_\textrm{ex}$. For large detunings, this cutoff appears at low momenta, so the assumption $\tau_\textrm{ph}\approx$~const is valid.

\begin{figure}[t]
  \includegraphics[width=0.75\linewidth]{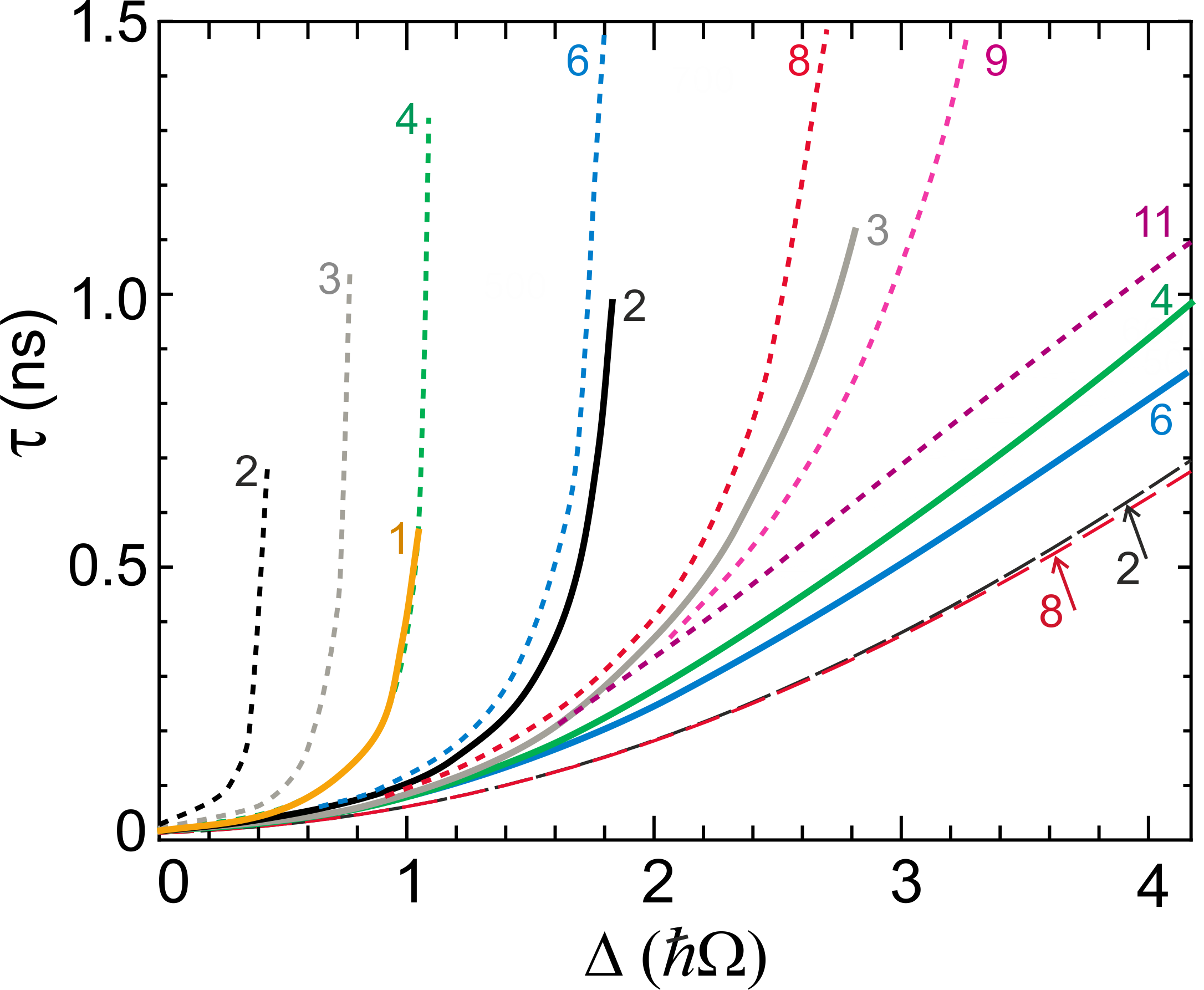}
  \linespread{1.0}\caption{(Color online)
  Integral polariton lifetime $\tau$ according to (\ref{tau_int}) dependent on the detuning, for $T=0$ (dashed), $T=10$~K (solid), $T=20$~K (dotted lines), for the densities $n = 10^{10}$ (yellow), $2\times10^{10}$ (black), $3\times10^{10}$ (gray), $4\times10^{10}$ (green), $6\times10^{10}$ (blue), $8\times10^{10}$ (red), $9\times10^{10}$ (magenta), and $1.1\times10^{11}$~cm$^{-2}$ (purple). For $T=0$, the lines for all densities almost coincide, so here we show the dependencies for $2\times10^{10}$ and $8\times10^{10}$~cm$^{-2}$ only. The slight difference of the two lines at high $\Delta$ is produced by using the Lennard-Jones potential as the exciton pair interaction. The rapid increase of the integral lifetime is due to the fact that with the growth of $\Delta$, the critical temperature $T_C$ is approached and the condensate fraction drops (at different $\Delta$ for different $n$). Since for smaller densities $T_C$ is reached earlier, a different number of lines is shown for 10~K and 20~K. In this figure $\hbar\Omega=7.2$~meV, $g = 2.5~\mu$eV~$\mu$m$^2$, $\tau_\textrm{ph} = 10$~ps, $\tau_\textrm{ex} = 0.5$~ns.
  }
\label{fig5_tau}
\end{figure}

The results of calculations according to (\ref{tau_int}) are shown in Fig.~\ref{fig5_tau} for $T=0$, 10, and 20~K, depending on $\Delta$. The slight deviation of the lines for the two densities at $T=0$ is the result of using the Lennard-Jones potential as the exciton-exciton interaction. If contact interaction with a fixed $g$ is used, the curves at $T=0$ for all $n$ coincide. 
For $T>0$, increasing temperature results in rapid growth of the lifetime with $\Delta$: a larger non-condensate fraction is longer-lived compared to the condensate particles, hence $\tau$ becomes larger.
The integral particle lifetime is a quantity of interest because the condensate and non-condensate fractions are mutually transforming into each other, and while the condensate lifetime can be short, the total lifetime of the system is much larger, and it is exactly the quantity that has to be compared with the relaxation time when discussing thermal equilibrium.
One can see that for moderate and large densities ($2\times10^{10}$~cm$^{-2}$ and higher), $\tau$ reaches nanoseconds, which is much larger than the expected relaxation time \cite{deng2006}.
This supports correctness of our original assumption of equilibrium at increased positive $\Delta$, justifying the developed theoretical approach. For $n\lesssim1\times10^{10}$~cm$^{-2}$ and for small or zero detunings, the theory is applicable for high-$Q$ microcavities which ensure larger $\tau_\textrm{ph}$.

\subsection{Critical temperature of condensation}

\begin{figure}[b]
  \includegraphics[width=0.90\linewidth]{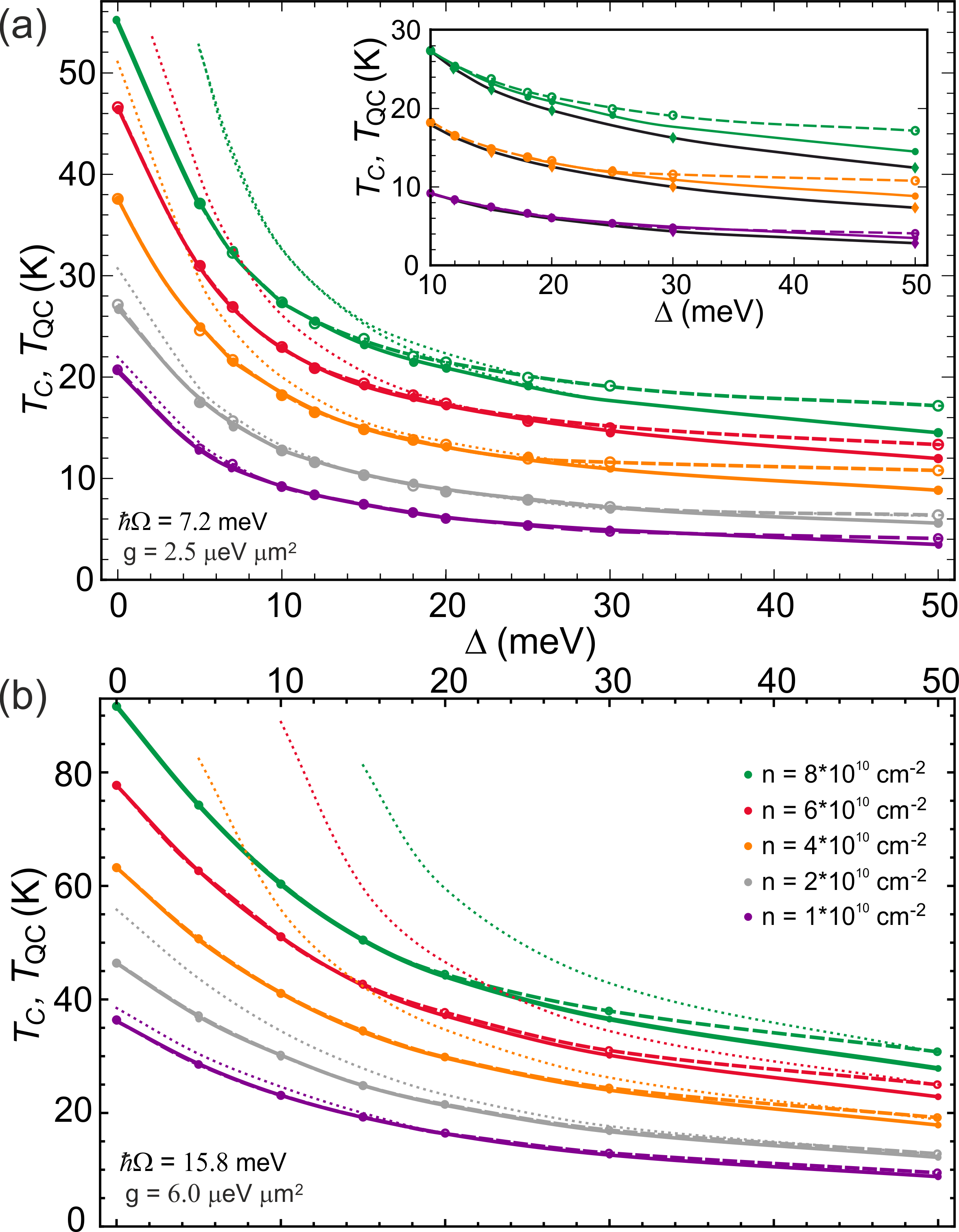}
  \linespread{1.0}\caption{(Color online) Critical temperature of Bose condensation $T_C$ (solid lines) and quasicondensation $T_{QC}$ (dashed lines) for different values of the total polariton density $n = 1$, $2$, $4$, $6$, $8\times10^{10}$~cm$^{-2}$ (bottom to top).
  {\bf a}, for $\hbar\Omega = 7.2$~meV, $g = 2.5~\mu$eV~$\mu$m$^2$. For $\Delta$ below $3\hbar\Omega$, $T_C$ coincides with $T_{QC}$ for all considered values of $n$.
  Inset: solid lines with closed circles show $T_C$ calculated with $L=100~\mu$m (same as in the main panel), diamonds for a larger system size $L=10^4~\mu$m, dashed lines (open circles) shows $T_{QC}$ as in the main panel, for $n=1$, $4$, $8\times10^{10}$~cm$^{-2}$.
  {\bf b}, for $\hbar\Omega = 15.8$~meV, $g = 6~\mu$eV~$\mu$m$^2$. Even for this case of increased interactions, the dependencies coincide for $n$ up to $3\times10^{10}$~cm$^{-2}$. Thin dotted lines of respective colors show $T_C$, $T_{QC}$ calculated with the background dark population $\tilde{n}$ taken into account in the assumption of thermal equilibrium, coinciding with the main lines for large delta and deviating considerably for $\Delta<2\hbar\Omega$ and large $n$.}
\label{fig6_Tc}
\end{figure}

The theory developed in Sec.~\ref{HFB_theory} allows to self-consistently define the critical temperature of Bose-Einstein condensation in a finite polariton system of the size $L$. For that, as in previous subsections, we solve the Eqs.~(\ref{int_system}) together with (\ref{bog_spectrum})---(\ref{n_p}) for each temperature, defining this way the dependence of the condensate density $n_0$ on $T$ for every value of the total density $n$ and the detuning $\Delta$ that we treat as external parameters. The critical temperature of transition is then found by extrapolating the dependence $n_0(T)\to0$.
Such calculations were performed for multiple values of $\hbar\Omega$ and $g$ in consideration. In  Fig.~\ref{fig6_Tc}, we show two cases of resulting dependencies of $T_C$ on $\Delta$ for each density $n$ (solid lines). Since the particles effective mass increases with $\Delta$ (see Fig.~\ref{fig2_m_B}b), the critical temperature drops from several tens of K at zero detuning to below 10--20~K (20--40~K) for small (large) Rabi splittings as $\Delta$ exceeds $2\hbar\Omega$. In this regime polaritons become exciton-like as their exciton fraction $X_0^2$ grows up to almost unity. These $T_C$ stay high compared to the temperatures of exciton condensation ($\sim0.1$~K \cite{butov}), since the polariton mass still stays orders of magnitude lower than $m_\textrm{ex}$ due to the maintained strong coupling to photons.

For these calculations, we assumed the system size $L=100~\mu$m and cut the integration in (\ref{int_system}) from below at $2\pi\hbar/L$. As the system size is increased, the Bogoliubov description starts to fail in accordance with the Hohenberg--Mermin--Wagner theorem \cite{pit_str}. Due to this reason, the HFB theory estimate of the condensate density $n_0$ (and hence $T_C$) at the increased $L$ starts to be incorrect. The inset of Fig.~\ref{fig6_Tc}a shows the comparison of calculations made with $L=100~\mu$m and the 100 times larger size $L=10^4~\mu$m: in the latter case, $T_C$ drops (at different $\Delta$ for different $n$). 
Сalculated in the HFB theory $n_0$ going down indicates the disappearance of the true Bose-Eistein condensate (BEC) in the system. In this case one needs to switch to the description in terms of the superfluid density $n_s$ instead of $n_0$, since the quasicondensate is still present even when the BEC is not. In Appendix~\ref{AppHD}, we show the mathematical way of stitching the HFB one-body density matrix $\rho_1({\bf r})\equiv\langle\hat{P}^\dag({\bf r})\hat{P}(0)\rangle$ obtained within the theory described in Sec.~\ref{HFB_theory}, with the hydrodynamic $\rho_1^\textrm{HD}({\bf r})$ used to describe the superfluid transition \cite{voronova_PRL}. This allows to switch from finding the critical temperature $T_C$ of BEC to defining the critical temperature $T_{QC}$ of quasi-condensation (as the temperature at which the local superfluidity and quasicondensate disappear). Using this stitching and considering the Bogoliubov excitations with the spectrum (\ref{bog_spectrum}) as non-interacting non-quasicondensate particles, one finds $T_{QC}$ as the temperature at which the quasicondensate density
\begin{equation}\label{n_qc}
n_{\rm qc} = n - \frac{1}{S}\sum_{\bf p \neq0}\hat{\alpha}^\dag_{\bf p}\hat{\alpha}_{\bf p} = n - \int \frac{1}{e^{E_p/T}-1}\frac{d{\bf p}}{(2\pi\hbar)^2}
\end{equation}
goes to zero: $n_{\rm qc}(T)\to0$. Effectively, when finding $T_{QC}$, Eq.~(\ref{n_qc}) replaces the first equation in (\ref{int_system}), whereas the polariton occupation number $n_p$ in the second line of (\ref{int_system}) is replaced by the Bose distribution of the excitations $\hat{\alpha}^\dag_{\bf p}\hat{\alpha}_{\bf p}$. The dependencies of $T_{QC}$ on $\Delta$ for different densities are plotted in Fig.~\ref{fig6_Tc} as dashed lines. One sees that at high densities $n$ and large detunings, $T_{QC}$ is higher than $T_C$ for the same parameters, as it should be, and the deviation is larger when the system size is increased (see the inset of Fig.~\ref{fig6_Tc}a). However, it is also evident that for not too large detunings ($\Delta \lesssim 3\hbar\Omega$), the HFB description works very well for all considered densities.

To finalize the analysis, we address the influence of the dark exciton population that we have added to our consideration according to (\ref{darkexcitons}), assuming $\tilde{g}=g$. The resulting $T_{C(QC)}$ dependencies on $\Delta$ are shown in Fig.~\ref{fig6_Tc} as dotted lines for each polariton density $n$. One sees that in the case of Fig.~\ref{fig6_Tc}a (weaker interactions, $g=2.5~\mu$eV~$\mu$m$^2$) $\tilde{n}$ has an influence on the critical temperature only in the region of small detunings and for large total densities, whereas for large $\Delta$, the tails of the dotted curves both for $T_C$ and $T_{QC}$ fully coincide with those calculated taking $\tilde{n}=0$. In the region of near-zero detunings the background particles shift the critical temperature to higher values. This rise of $T_C$ corresponds to the lowering of the renormalized effective mass $m_B$ with respect to $m_\textrm{LP}$ which is shown in Fig.~\ref{fig2_m_B}a by dotted lines. When considering the case of increased interaction (see Fig.~\ref{fig6_Tc}b for $g=6~\mu$eV~$\mu$m$^2$ and the red dotted line in Fig.~\ref{fig2_m_B}a), the situation is the same for small densities $n\sim1$--$2\times10^{10}$~cm$^{-2}$, whereas for large densities $n\gtrsim4\times10^{10}$~cm$^{-2}$ the deviation at low detunings is very large. However for detunings $\Delta > 2\hbar\Omega$ their influence diminishes similarly to the case of Fig.~\ref{fig6_Tc}a.

\section{Applicability}

\begin{figure}[b]
  \centering
  \includegraphics[width=0.85\linewidth]{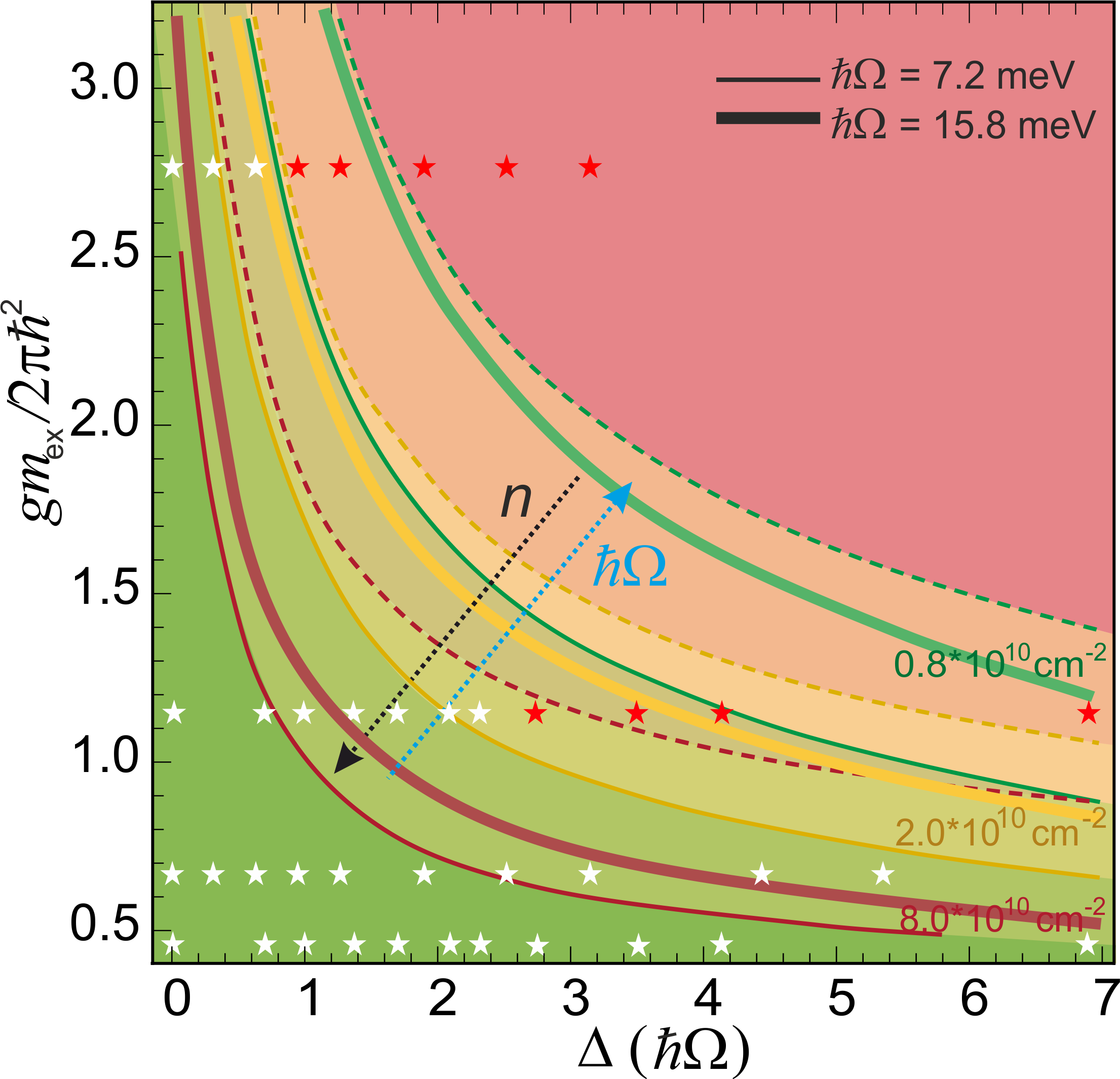}
  \linespread{1.0}\caption{(Color online)
  Diagram of applicability of the Hartree--Fock--Bogoliubov description in the domain of dimensionless detuning and exciton interaction strength (here $m_\textrm{ex} = 0.22m_0$). Solid lines indicate the boundary $\nu^\prime_{T=0}\sim0.1$ for the densities as marked, with the dark thin lines corresponding to $\hbar\Omega=7.2$~meV and lighter thick lines to $\hbar\Omega=15.8$~meV. Dashed lines represent the boundary defined by a weaker condition $\nu^\prime_{T=0}\sim0.2$ for $\hbar\Omega = 7.2$~meV. The stars mark the studied parameters combinations. Red markers indicate the parameters at which the hydrodynamic description ($T_{QC}$) noticeably differs from the HFB description ($T_C$).
  }
\label{fig7_applicability}
\end{figure}

Validity of the developed HFB description is limited by two factors. First condition which was addressed in Sec.~\ref{sec_lifetime} is that the lifetimes should large enough to assume equilibrium. The second condition is that even with the stitching to the hydrodynamic description one needs to make sure that the condensate depletion of the polariton gas at $T=0$ is small: $\nu^\prime_{T=0}\ll1$, where $\nu^\prime=n^\prime/n$ is calculated in the Bogoliubov approach (see Appendix~\ref{AppBog}). The hydrodynamic description is expected to work fairly well up to $\nu^\prime_{T=0}\sim0.5$. Depending on the interaction strength and the detuning, this condition is violated at different total densities $n$. Using (\ref{depletion_bog}), we plot an applicability diagram in terms of the dimensionless exciton interaction constant $gm_\textrm{ex}/2\pi\hbar^2$ and the dimensionless detuning $\Delta/\hbar\Omega$, for different $\hbar\Omega$ and $n$ (see Fig.~\ref{fig7_applicability}). The diagram shows that the theory developed in this paper works very well for the large range of densities, interactions, and detunings (the green-shaded area of the diagram shows the parameters at which the pure HFB description is applicable, while the red-shaded area indicates the region where the stitching with hydrodynamics is required). It is also worth noting that for higher Rabi splittings the theory works better. We indicate in Fig.~\ref{fig7_applicability} the parameters combinations that were addressed within this study, and the white color of those markers show that the HFB description has proved to be working well ({\it i.e.} $\nu^\prime_{T=0}\sim0.1$), whereas the red color is chosen for the parameters where we had to replace the true condensate $n_0$ with the quasicondensate density $n_{\rm qc}$ as described above.

\section{Conclusions}

In this work, we have derived the finite-temperature corrections and modifications produced by the momentum-dependent interactions within the Hartree--Fock--Bogoliubov theory applied to a system of exciton-polaritons at positive detunings. The developed theory yields renormalizations of the bare particle spectrum and the particle effective mass, both at $T=0$ and $T>0$, resulting in the shifts of the chemical potential and the spectrum of collective excitations of the polariton system. Notably, the modifications that we obtain within the equilibrium theory are shown to produce the deviations of the sound velocity from the standard Bogoliubov value, that are routinely observed in exciton-polariton experiments and which were previously attributed to effects of dissipation. In order to correctly address systems of an increased size, we provide the stitching with the hydrodynamic approach employed to describe the superfluid transition. We show that for large detunings, the integral lifetime of polaritons grows up to the nanosecond scale even for microcavities where the photon lifetimes are not very high ($\sim10$~ps), whereas the critical temperature of transition to the macroscopically coherent state in thermal equilibrium still stays as high as 10--20~K, dependent on the density. This suggests such systems with ``shallow polariton well'' (large positive detunings), where polaritons become extremely exciton-like yet staying strongly coupled to photons, to be natural candidates for experimental observation of the long-living thermally-equilibrium polariton systems with BEC.
Finally, the treatment of the background particles ({\it e.g.} dark and incoherent excitons which affect the system by interacting with the polariton's excitonic fraction) is included, with the assumption of thermal equilibrium with the semiconductor lattice. We leave the treatment of the non-equilibrium background particles to future work.

\begin{acknowledgments}
The authors are thankful to A. Semenov for discussions. A.M.G., Yu.E.L. and N.S.V. acknowledge the financial support of Russian Foundation for Basic Research within the joint DFG/RFBR project No.~21--52--12038. I.L.K. and the research on quantum hydrodynamics is supported by the RFBR grant No.~19--02--00793. Yu.E.L. is supported by the Program of Basic Research of the Higher School of Economics.
\end{acknowledgments}

\appendix
\section{Polariton modification of the Bogoliubov theory at $T=0$.}\label{AppBog}
For completeness of the analysis, here we present the Bogoliubov theory \cite{bogoliubov,AGD} for Bose-condensed polaritons at $T=0$, when the condensate density $n_0$ is close to the total density $n$. Starting from the second-quantized Hamiltonian of the polariton system in momentum basis, $\hat{H} = \hat{H}_0 + \hat{U}_\textrm{LP}$, where the first term describes the ideal gas of lower polaritons $\hat{H}_0 = \sum_{\bf p} \varepsilon_p \hat{P}_{\bf p}^\dag\hat{P}_{\bf p}$, and the interaction is defined by the exciton-exciton pair interaction
\begin{equation}\label{U}
\hat{U}_\textrm{LP} = \frac{1}{2S}\!\sum\limits_{{\bf p}_1+{\bf p}_2 = {\bf p}_3+{\bf p}_4}\!\!\!\!\!\! U({\bf p}_1-{\bf p}_3)\hat{Q}_{{\bf p}_1}^\dag\hat{Q}_{{\bf p}_2}^\dag \hat{Q}_{{\bf p}_3}\hat{Q}_{{\bf p}_4},
\end{equation}
we separate the condensate fraction in the exciton operator,
$\hat{Q}_{\bf p} = \delta_{\bf p0}X_0\sqrt{n_0} + (1 - \delta_{\bf p0})X_p\hat{P}_{\bf p}$. As long as the condensate depletion is small compared to the condensate density $n-n_0\ll n_0$, the condensate contribution is dominant: $\hat{Q}_{{\bf p}\neq0}=X_0\sqrt{n_0\,\mathcal{O}[(n-n_0)/n_0]} \ll\hat{Q}_0= X_0\sqrt{n_0}$. Keeping only the first non-vanishing (quadratic) terms with respect to $\hat{P}_{\bf p}$ in (\ref{U}), one gets
\begin{multline}\label{U_bog}
\hat{U}_\textrm{LP} = \frac{S}{2}\,gn_0^2X_0^4 \\
+ \frac{n_0}{2}\sum\limits_{{\bf p}\neq0} \!X_0^2X_p^2\left\{\bigl[g + U({\bf p})\bigr]\!\left(\hat{P}_{\bf p}^\dag \hat{P}_{\bf p} + \hat{P}_{\bf -p}^\dag\hat{P}_{\bf -p}\!\right) \right. \\
\left. + U({\bf p})\hat{P}_{\bf p}\hat{P}_{\bf -p} + U({\bf p})\hat{P}_{\bf p}^\dag \hat{P}_{\bf -p}^\dag\right\}.
\end{multline}
As we choose to fix the total density $n$, the substitution $n_0 = n - \frac{1}{2S}\sum\limits_{{\bf p}\neq0} \left(\hat{P}_{\bf p}^\dag \hat{P}_{\bf p} + \hat{P}_{\bf -p}^\dag\hat{P}_{\bf -p}\!\right)$ yields for the Hamiltonian of the system:
\begin{multline}\label{H_bog}
\hat{H} = \sum\limits_{\bf p}\varepsilon_p^{B(T=0)}\hat{P}_{\bf p}^\dag \hat{P}_{\bf p} + \frac{n}{2}\sum\limits_{{\bf p}\neq0}X_0^2X_p^2 U({\bf p})\left(\hat{P}_{\bf p}^\dag \hat{P}_{\bf p} \right. \\
\left. + \hat{P}_{\bf -p}^\dag\hat{P}_{\bf -p} + \hat{P}_{\bf p}\hat{P}_{\bf -p} + \hat{P}_{\bf p}^\dag\hat{P}_{\bf -p}^\dag\right),
\end{multline}
where the null-particle (constant) terms are omitted, and
\begin{eqnarray}
\varepsilon_p^{B(T=0)} &=& \varepsilon_p + gnX_0^2\left(X_p^2-X_0^2\right) \label{eps_bog_T0}\\
&\approx&
\left\{
\begin{array}{lc}
  p^2/2m_B^{(T=0)}, & p\ll\tilde{p} \\
  E_0 + gnX_0^2(1-X_0^2) + p^2/2m_\textrm{ex}, & p\gg\tilde{p}
\end{array}
\right.\nonumber
\end{eqnarray}
is the renormalized particle spectrum within the polariton Bogoliubov theory, with
\begin{equation}\label{m_B_0}
\frac{1}{m_B^{(T=0)}} = \frac{1}{m_\textrm{LP}}\left(1 + \frac{2gnX_0^4}{\sqrt{\Delta^2 + (\hbar\Omega)^2}}\right).
\end{equation}
Diagonalizing the Hamiltonian (\ref{H_bog}) by the standard Bogoliubov transformation,
one gets the spectrum of the shape (\ref{bog_spectrum}) and the Bogoliubov coefficients (\ref{uv_bog}), with the replacements $\varepsilon_p^B\rightarrow \varepsilon_p^{B(T=0)}$ and $\mu_p \rightarrow U({\bf p})nX_0^2X_p^2$ (compared to (\ref{mu_p}), here the condensate density $n_0$ is replaced with the total density $n$). Thus even at $T=0$, the excitation spectrum contains an extra momentum dependence compared to the regularly used Bogoliubov dispersion. The diagonalized form of the polariton Hamiltonian (\ref{H_bog}) allows to find the occupation number of polaritons in the Bogoliubov approximation at $T=0$,
\begin{equation}
n_p = v_p^2,
\end{equation}
and the corresponding condensate depletion:
\begin{multline}\label{depletion_bog}
\nu^\prime_{T=0} \equiv \frac{n-n_0}{n}
= \frac{1}{nS}\sum\limits_{{\bf p}\neq0}v_p^2 \\
=\! \frac{1}{2n} \!\int\!\! \left(\!\sqrt{1+\frac{\mu_p^2}{(\varepsilon_p^{B(T=0)}\!\!+\mu_p)^2-\mu_p^2}}-1\!\right)\!\! \frac{d{\bf p}}{(2\pi\hbar)^2}.
\end{multline}
The obtained expression (\ref{depletion_bog}) contains the full renormalized polariton spectrum (\ref{eps_bog_T0}), thus taking into account the states corresponding to the high-energy reservoir, and the shift of the polariton effective mass and the chemical potential due to the dependence of interactions on momentum.
In is worth noting that the existence of the characteristic momentum $\tilde{p}$ allows to approximately split the non-condensate fraction into the polariton and exciton constituents, $\nu^\prime_{T=0} = \nu^\prime_\textrm{LP} + \nu^\prime_\textrm{ex}$, with
\begin{multline}
\nu^\prime_\textrm{LP}\approx\!\! \int\limits_0^{\tilde{p}}\! \left[\sqrt{1+\frac{(gnX_0^4)^2}{(p^2/2m_B+ gnX_0^4)^2 - (gnX_0^4)^2}} \right.\\
\Bigl. - 1 \Bigr]\!\frac{pdp}{4\pi\hbar^2 n} \nonumber
\end{multline}
being the expression regularly used in the literature as the polariton condensate depletion (however with $m_\textrm{LP}$ instead of $m_B^{(T=0)}$), and
\begin{multline}
\nu^\prime_\textrm{ex} \approx\!\!\!\int\limits_{\tilde{p}\to0}^\infty \!\!\! \left\{\!\sqrt{1+\frac{\mu_\textrm{ex}^2}{[E_0 + \mu_\textrm{ex}(2-X_0^2) + p^2/2m_\textrm{ex}]^2 - \mu_\textrm{ex}^2}} \right.\\
\Bigl. - 1\Bigr\}\! \frac{pdp}{4\pi\hbar^2 n}, \nonumber
\end{multline}
corresponding to the exciton-like part of the polariton spectrum (here we introduced the notation $\mu_\textrm{ex}=gnX_0^2$).

\section{Non-condensate anomalous average} \label{AppM}

\begin{figure}[h]
  \centering
  \includegraphics[width=1.0\linewidth]{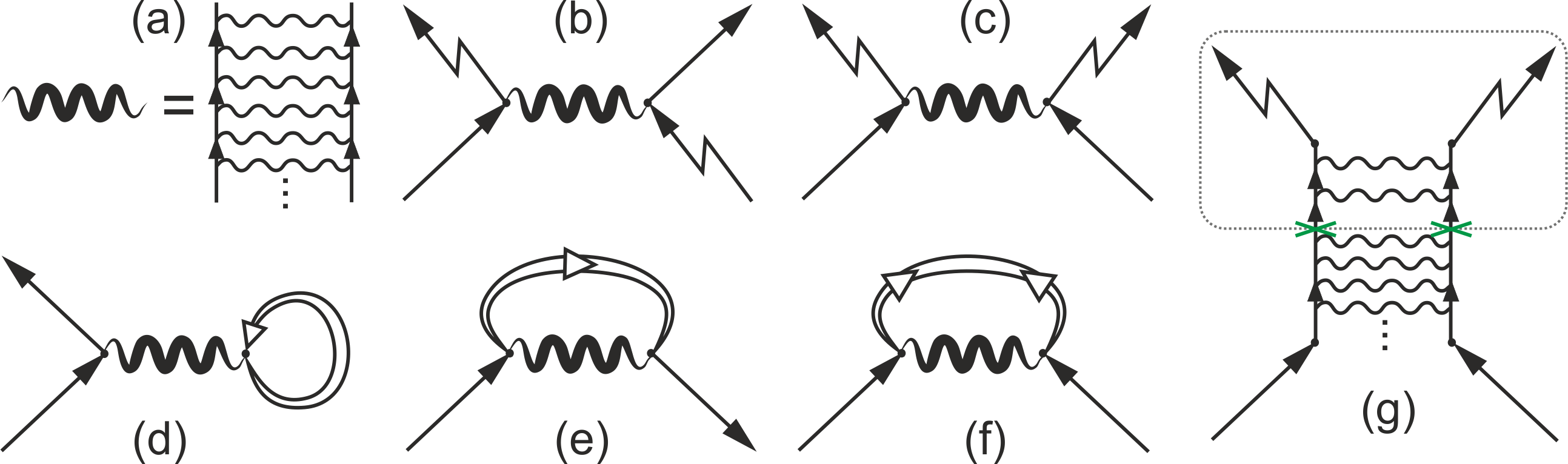}
  \linespread{1.0}\caption{{\bf a}, Graphic representation of the pair interaction as a vertical ladder diagram. Diagrams: {\bf b,c}, normal and anomalous condensate averages, {\bf d}, non-condensate exciton density $n^\prime_Q$, {\bf e}, non-condensate one-body density matrix $\rho^\prime_{1Q}({\bf r},{\bf r}^\prime)$, {\bf f}, non-condensate anomalous average $m^\prime_Q({\bf r},{\bf r}^\prime)$. {\bf g}, A more detailed drawing of the diagram {\bf c}. Сrosses show one of possible cuts of the diagram, after which the part in the dotted rectangle is partly contained in the anomalous Green's function in the panel {\bf f}.}
\label{fig8_diagrams}
\end{figure}

In the Beliaev formalism \cite{beliaev} that we use, the pair interaction is substituted by a ladder diagram shown in Fig.~\ref{fig8_diagrams}a \cite{schick,LozYud}. It is important to note that since the Beliaev ladder contains directed arrows pointing up, the incoming lines are always at the bottom and the outgoing lines are always at the upper side of the ladder.
Here we present the diagrams corresponding to the terms in the r.h.s. of (\ref{3oper_int}) in the Hatree--Fock--Bogoliubov approach for the case $T>0$. In particular, the condensate normal and anomalous average terms, both equal to $X_0^2n_0$, correspond to Fig.~\ref{fig8_diagrams}b and c, respectively. The diagrams corresponding to the non-condensate diagonal and off-diagonal densities $n^\prime_Q$ and $\rho_ {1Q}^\prime({\bf r},{\bf r}^\prime)$ which are given in the two top lines in (\ref{n'}) are shown in the panels (d) and (e). Finally, the non-condensate anomalous average $m^\prime_Q({\bf r},{\bf r}^\prime)$, given by the third line in (\ref{n'}), corresponds to the diagram in the panel (f). It is clear from direct diagrammatic calculation that the non-condensate anomalous average (f) is already partly included in the condensate anomalous term (c) which is shown in more detail in Fig.~\ref{fig8_diagrams}g (see the part enclosed in the dotted rectangle). Therefore to avoid double counting, the dotted rectangle in (g) should be subtracted from the diagram (f). The remaining after subtraction parts of (f) have a higher degree of smallness compared to all the other diagrams in the panels (b)--(e) and their omission does not produce error. Therefore one can discard the term corresponding to (f) in the equations: $m^\prime_Q({\bf r},{\bf r}^\prime)=0$.

\section{Stitching with the hydrodynamic description} \label{AppHD}

As described in the main text, to substitute the BEC density $n_0$ with the quasicondensate density $n_{\rm qc}$, one needs to perform the stitching of the HFB description with the hydrodynamic (HD) description. The non-condensate occupation number $n_p$ calculated with the HFB approach is given by (\ref{n_p}), while in HD it is given by the Bose--Einstein distribution $n_p^{\rm (qc)} = 1/[\exp(E_p/T)-1]$, in the fair assumption $|n_{\rm qc}-n_s|\ll n_s$ ($n_s$ being the superfluid density). The quantum hydrodynamic approach, where the one-body density matrix was obtained in the long-wavelength limit, {\it i.e.} at large distances $r\sim L$, was developed in Ref.~\cite{voronova_PRL,boronat} to describe superfluidity and the Berezinskii--Kosterlitz--Thouless crossover in a finite system of 2D excitons. For polaritons with the dispersion (\ref{epsilon_B}), assuming for simplicity that there are no vortices in the system, the expression obtained in \cite{voronova_PRL} can be rewritten as
\begin{equation}\label{rho1_hd}
  \frac{\rho_1^\textrm{HD}({\bf r})}{n} \!\sim \exp\!\!\left[\frac{-1}{\nu_sN}\! \sum\limits_{\bf p\neq0}\!\frac{\kappa_pE_p}{4\varepsilon_p^B} \!\left(\!1 \!-\! \cos\frac{{\bf p}\!\cdot\!{\bf r}}{\hbar}\right)\!\!\!\left(\!1 \!+\! \frac{2}{e^\frac{E_p}{T}-1}\!\right)\!\!\right]\!\!,
\end{equation}
where $ \nu_s = n_s/n$ is the superfluid fraction, $N=nS$ is the total number of particles, and the spectrum of excitations $E_p$ is given by (\ref{bog_spectrum}) with $n_s$ instead of $n_0$. The ultraviolet cutoff is chosen in such a way that $\kappa_p=1$ at $p\to0$ and $\kappa_p=0$ at $p\to\infty$.

The one-body density matrix for the polariton field in HFB approximation (using (\ref{bog_subs}))
\begin{equation}
 \rho_1({\bf r})\equiv\langle\hat{P}^\dag({\bf r})\hat{P}(0)\rangle = n - \frac{1}{S}\sum \limits_{\bf p\neq0}n_p\left(1 - e^{-\frac{i}{\hbar}{\bf p}\cdot{\bf r}}\right),
\end{equation}
with $n_p = v_p^2 + (2v_p^2+1)/(e^{E_p/T}-1)$, after some algebra can be brought to the form
\begin{multline}\label{rho1}
\frac{\rho_1({\bf r})}{n} = \frac{1}{N}\sum\limits_{\bf p\neq0} \frac{\cos({\bf p}\!\cdot\!{\bf r}/\hbar)}{e^{E_p/T}-1} \\
+ \nu_{\rm qc} -\frac{1}{N}\sum\limits_{\bf p\neq0}\frac{(E_p-\varepsilon_p^B)^2}{4 E_p\varepsilon_p^B} \!\left(\!1 \!-\! \cos\frac{{\bf p}\!\cdot\!{\bf r}}{\hbar}\right)\!\!\left(\!1 \!+\! \frac{2}{e^{E_p/T}-1}\!\right)
\end{multline}
with the quasicondensate fraction
\begin{equation}\label{nu_q}
  \nu_{\rm qc} = \frac{n_{\rm qc}}{n} = 1 - \frac{1}{N}\sum\limits_{\bf p\neq0}\frac{1}{e^{E_p/T}-1}.
\end{equation}
The first term in (\ref{rho1}) represents the amplitude of the algebraic order (as long as the quasicondensate is present in the system). The rest of (\ref{rho1}) can be consistently stitched to $\rho_1^\textrm{HD}({\bf r})$ given by (\ref{rho1_hd}), making use of the following transformations.

\noindent In (\ref{rho1_hd}),
\begin{itemize}
\item[(i)] the short-distance cutoff is taken in the shape
\vspace{-1em}
$$\kappa_p = \left(1-\frac{\varepsilon_p^B}{E_p}\right)^2;$$
\item[(ii)] the proportionality coefficient in front of the exponent is taken equal $\nu_{\rm qc}$.
\end{itemize}
In (\ref{rho1}),
\begin{itemize}
\item[(iii)] $\mu_p=gn_0X_0^2X_p^2$ is replaced by $gn_sX_0^2X_p^2$;
\item[(iv)] $\displaystyle \nu_{\rm qc} - \frac{1}{N}\sum\limits_{\bf p\neq0} \dots = \nu_{\rm qc} \Bigl(1 - \frac{1}{\nu_{\rm qc}N}\sum\limits_{\bf p\neq0}\dots + \dots\Bigr)$
     is replaced by $\displaystyle \nu_{\rm qc}\exp\Bigl[-\frac{1}{\nu_{\rm qc}N}\sum\limits_{\bf p\neq0}\dots\Bigr]$;
\item[(v)] if the vortex renormalizations inside $\exp[\dots]$ are required, one needs to replace $n_s$ with the superfluid density renormalized by vortex pairs $\tilde{n}_s$, according to Kosterlitz \cite{kosterlitz}.  To account for free vortices, the whole expression is multiplied by $\exp(-r/\xi_+)$, where $\xi_+$ is the distance between free vortices \cite{voronova_PRL}.
\end{itemize}

With these transformations, one gets the coinciding expressions except the first term in (\ref{rho1}). The stitched one-body density matrix, assuming $\nu_q\approx\nu_s$, has the form:
\begin{multline}\label{stitched}
  \frac{\rho_1({\bf r})}{n} = \frac{1}{N}\sum\limits_{\bf p\neq0} \frac{\cos({\bf p}\!\cdot\!{\bf r}/\hbar)}{e^{E_p/T}-1} \\
  + \nu_{\rm qc}\!\exp\!\!\left[\frac{-1}{\nu_{\rm qc}N}\!\sum\limits_{\bf p\neq0}\!\!\frac{(E_p-\varepsilon_p^B)^2}{4 E_p\varepsilon_p^B} \!\left(\!1 \!-\! \cos\frac{{\bf p}\!\cdot\!{\bf r}}{\hbar}\right)\!\!\!\left(\!1 \!+\! \frac{2}{e^\frac{E_p}{T}-1}\!\right)\!\!\right]\!\!.
\end{multline}
The stitched occupation number $n_p$ is then obtained by making the Fourier transform of (\ref{stitched}).


\begin{thebibliography}{99}
\bibitem{keeling2006}
J.~Keeling, {\it Response functions and superfluid density in a weakly interacting Bose gas
with nonquadratic dispersion}, Phys. Rev. B {\bf 74}, 155325 (2006).

\bibitem{semenov}
A. Semenov and Yu. Lozovik, {\it On the superfluid properties of a polaritonic system}, EPL {\bf 78}, 67005 (2007).

\bibitem{Microcavities}
A.~V.~Kavokin, J.~J.~Baumberg, G.~Malpuech, F.~P.~Laussy, {\it Microcavities} 2nd Ed., Oxford University Press (2017).

\bibitem{RMP2010}
H. Deng, H. Haug, and Y. Yamamoto, {\it Exciton-polariton Bose-Einstein condensation}, Rev. Mod. Phys. {\bf 82}, 1489--1537 (2010).

\bibitem{QFL}
I. Carusotto and C. Ciuti, {\it Quantum fluids of light}, Rev. Mod. Phys. {\bf 85}, 299 (2013).

\bibitem{wouters2007}
M. Wouters and I. Carusotto, {\it Excitations in a nonequilibrium Bose-Einstein condensate of exciton polaritons}, Phys. Rev. Lett. {\bf 99}, 140402 (2007).

\bibitem{berloff}
J. Keeling and N. G. Berloff, {\it Spontaneous Rotating Vortex Lattices in a Pumped Decaying Condensate}, Phys. Rev. Lett. {\bf 100}, 250401 (2008).

\bibitem{manni}
F. Manni, K. G. Lagoudakis, T. C. H. Liew, R. Andr\'{e}, and B. Deveaud-Pl\'{e}dran, {\it Spontaneous Pattern Formation in a Polariton Condensate}, Phys. Rev. Lett. {\bf 107}, 106401 (2011).

\bibitem{haug14}
H. Haug, T. D. Doan and D. B. Tran Thoai, {\it Quantum kinetic derivation of the nonequilibrium Gross-Pitaevskii equation for nonresonant excitation of microcavity polaritons}, Phys. Rev. B {\bf 89}, 155302 (2014).

\bibitem{hybridBGP}
D. D. Solnyshkov, H. Ter\c{c}as, K. Dini, and G. Malpuech, {\it Hybrid Boltzmann–Gross-Pitaevskii theory of Bose-Einstein condensation and superfluidity
in open driven-dissipative systems}, Phys. Rev. A {\bf 89}, 033626 (2014).

\bibitem{haug2020}
T. D. Doan, D. B. Tran Thoai, and H. Haug, {\it Kinetics and luminescence of the excitations of a nonequilibrium polariton condensate}, Phys. Rev. B {\bf 102}, 165126 (2020).

\bibitem{yamamoto2012}
T. Byrnes, T. Horikiri, N. Ishida, M., Fraser, and Y. Yamamoto, {\it The negative
Bogoliubov dispersion in exciton-polariton condensates}, Phys. Rev. B. {\bf 85}, 075130 (2012).

\bibitem{ostrovskaya2014}
L. A. Smirnov, D. A. Smirnova, E. A. Ostrovskaya, and Yu. S. Kivshar, {\it Dynamics and stability of dark solitons in exciton-polariton condensates}, Phys. Rev. B {\bf 89}, 235310 (2014).

\bibitem{utsunomiya}
S. Utsunomiya, L. Tian, G. Roumpos, C.~W. Lai, N. Kumada, T. Fujisawa, M. Kuwata-Gonokami, A. L\"{o}ffler, S. H\"{o}fling, A. Forchel and Y. Yamamoto, {\it Observation of Bogoliubov excitations in exciton-polariton condensates}, Nature Phys. {\bf 4}, 700--705 (2008).

\bibitem{kohnle}
V. Kohnle, Y. L\'{e}ger, M. Wouters, M. Richard, M. T. Portella-Oberli, and B. Deveaud-Pl\'{e}dran, {\it From single particle to superfluid excitations in a dissipative polariton gas}, Phys. Rev. Lett. {\bf 106}, 255302 (2011).

\bibitem{pieczarka2015}
M. Pieczarka, M. Syperek, {\L}. Dusanowski, J. Misiewicz, F. Langer, A. Forchel, M. Kamp, C. Schneider, S. H\"{o}fling, A. Kavokin, and G. S\c{e}k, {\it Ghost Branch Photoluminescence From a Polariton Fluid Under Nonresonant Excitation}, Phys. Rev. Lett. {\bf 115}, 186401 (2015).

\bibitem{stepanov}
P. Stepanov, I. Amelio, J.-G. Rousset, J. Bloch, A. Lema\^{\i}tre, A. Amo, A. Minguzzi, I. Carusotto, and M. Richard, {\it Dispersion relation of the collective excitations in a resonantly driven polariton fluid}, Nat. Commun. {\bf 10}, 3869
(2019).

\bibitem{ballarini2020}
D. Ballarini, D. Caputo, G. Dagvadorj, R. Juggins, M. De Giorgi, L. Dominici, K.West, L. N. Pfeiffer, G. Gigli, M. H. Szyma\'{n}ska, and D. Sanvitto, {\it Directional Goldstone waves in polariton condensates close to equilibrium}, Nat. Commun. 11, 217 (2020).

\bibitem{pieczarka2020}
M. Pieczarka, E. Estrecho, M. Boozarjmehr, O. Bleu, M. Mark, K. West, L. N. Pfeiffer, D.W. Snoke, J. Levinsen, M. M. Parish, A. G. Truscott, and E. A. Ostrovskaya, {\it Observation of quantum depletion in a non-equilibrium exciton-
polariton condensate}, Nat. Commun. {\bf 11}, 429 (2020).

\bibitem{estrecho2021}
E. Estrecho, M. Pieczarka, M. Wurdack, M. Steger, K. West, L. N. Pfeiffer, D.W. Snoke, A. G. Truscott, and E. A. Ostrovskaya, {\it Low-Energy Collective Oscillations and Bogoliubov Sound in an Exciton-Polariton Condensate}, Phys. Rev. Lett. {\bf 126}, 075301 (2021).

\bibitem{pit_str}
L. Pitaevskii and S. Stringari, {\it Bose-Einstein Condensation and Superfluidity}, Oxford: Oxford University Press, 2016.

\bibitem{sarchi}
D. Sarchi and V. Savona, {\it Spectrum and thermal fluctuations of a microcavity polariton Bose-Einstein condensate}, Phys. Rev. B {\bf 77}, 045304 (2008).

\bibitem{voronova_PRL}
N. S. Voronova, I. L. Kurbakov, and Yu. E. Lozovik, {\it Bose Condensation of Long-Living Direct Excitons in an Off-Resonant Cavity}, Phys. Rev. Lett. {\bf 121}, 235702 (2018).

\bibitem{boronat}
Yu. E. Lozovik, I. L. Kurbakov, G. E. Astrakharchik, and J. Boronat, {\it Estimation of the condensate fraction from the static structure factor}, Phys. Rev. B {\bf 103}, 094511 (2021).

\bibitem{deng2006}
H. Deng, D. Press, S. G\"{o}tzinger, G.S. Solomon, R. Hey, K.H. Ploog, and Y. Yamamoto, {\it Quantum Degenerate Exciton-Polaritons in Thermal Equilibrium}, Phys. Rev. Lett. {\bf 97}, 146402 (2006).

\bibitem{butov}
A. A. High, J. R. Leonard, A. T. Hammack, M. M. Fogler, L. V. Butov, A. V. Kavokin, K. L. Campman, and A. C. Gossard, {\it Spontaneous coherence in a cold exciton gas}, Nature (London) 483, 584 (2012).

\bibitem{yukalov}
V. I. Yukalov, {\it Basics of Bose-Einstein Condensation}, Phys. Part. Nucl. {\bf 42}, 460 (2011).

\bibitem{griffin}
A. Griffin, {\it Conserving and gapless approximations for an inhomogeneous Bose gas at finite temperatures}, Phys. Rev. B {\bf 53}(14), 9341 (1996).

\bibitem{darkex}
More precisely, we assume that the self-energies responsible for spin relaxation of excitons [$\hbar/\Sigma_{\rm spin}^{\rm relax}\sim40$--$60$~ps, see, {\it e.g.}, P.~Le~Jeune, X.~Marie, T.~Amand, F.~Romstad, F.~Perez, J.~Barrau, and M.~Brousseau, {\it Spin-dependent exciton-exciton interactions in quantum wells}, Phys. Rev. B {\bf 58}, 4853 (1998); D.~W.~Snoke, W.~W.~R\"{u}hle, K.~K\"{o}hler, K.~Ploog, {\it Spin flip of excitons in GaAs quantum wells}, Phys. Rev. B {\bf 55}, 13789 (1997)] are much smaller than the HFB self-energies that are taken into account here ($\hbar/gnX_0^2\sim1$~ps) and at the same time much larger than the self-energies of the particle decay ($\tau\sim1$~ns). In this case, the system acquires one joint chemical potential, while the mutual transformations of bright and dark excitons are negligible.

\bibitem{beliaev}
S. Beliaev, {\it Energy spectrum of a non-ideal Bose gas}, Sov. Phys. JETP {\bf 7}, 299 (1958).

\bibitem{meera}
G. Li , O. Bleu , M. M. Parish , and J. Levinsen, {\it Enhanced Scattering between Electrons and Exciton-Polaritons in a Microcavity}, Phys. Rev. Lett. {\bf 126}, 197401 (2021).

\bibitem{brightex}
Calculating the occupation of the polaritonic region $p<\tilde{p}$ for non-condensed particles populating the second bright branch $\sigma=2$, with $\varepsilon_p$ given by (\ref{epsilon_p}), for all considered $T$ and $\Delta$ yields $\tilde{n}_2(p<\tilde{p})$ to be at least three orders of magnitude smaller than $n + \tilde{n}_2(p>\tilde{p})$, where $n$ is the density of Bose condensing polaritons ($\sigma=1$) and $\tilde{n}_2(p>\tilde{p})$ is calculated with the dipersion $E_0+p^2/2m_\textrm{ex}$ as given in (\ref{darkexcitons}).

\bibitem{popov}
V.N. Popov, {\it Functional Integrals and Collective Modes},  Chap. 6, Cambridge
University Press, New York (1987).

\bibitem{estrecho_TF}
E. Estrecho, T. Gao, N. Bobrovska, D. Comber-Todd, M. D. Fraser, M. Steger, K. West, L. N. Pfeiffer, J. Levinsen, M. M. Parish, T. C. H. Liew, M. Matuszewski, D. W. Snoke, A. G. Truscott, and E. A. Ostrovskaya, {\it Direct measurement of polariton-polariton interaction strength in the Thomas-Fermi regime of exciton-polariton condensation}, Phys. Rev. B {\bf 100}, 035306 (2019).

\bibitem{bleu}
O. Bleu, G. Li, J. Levinsen, and M. M. Parish, {\it Polariton interactions in microcavities with atomically thin semiconductor layers}, Phys. Rev. Res. {\bf 2}, 043185 (2020).

\bibitem{blueshifts}
T.~Yagafarov, D.~Sannikov, A.~Zasedatelev, K.~Georgiou, A.~Baranikov, O.~Kyriienko, I.~Shelykh, L.~Gai, Z.~Shen, D.~Lidzey, and P.~Lagoudakis, {\it Mechanisms of blueshifts in organic polariton condensates}, Commun. Phys. {\bf 3}, 18 (2020).

\bibitem{byrnes_dip}
T. Byrnes, G. V. Kolmakov, R. Ya. Kezerashvili, and Y. Yamamoto, {\it Effective interaction and condensation of dipolaritons in coupled quantum wells}, Phys. Rev. B {\bf 90}, 125314 (2014).

\bibitem{deng_PNAS}
H. Deng, G. Weihs, D. Snoke, J. Bloch, and Y. Yamamoto, {\it Polariton lasing vs. photon lasing in a semiconductor microcavity}, PNAS {\bf 100}(26), 15318-15323 (2003).

\bibitem{bogoliubov}
N. Bogoliubov, {\it On the theory of superfluidity}, Acad. Sci. USSR. J. Phys. {\bf 11}, 23–32 (1947).

\bibitem{AGD}
A. A. Abrikosov, L. P. Gorkov, I. E. Dzyaloshinski, {\it Methods of Quantum Field Theory in Statistical Physics}, Englewood Cliffs, N.J.: Prentice-Hall (1963).

\bibitem{schick}
M. Schick, {\it Two-Dimensional System of Hard-Core Bosons}, Phys. Rev. A {\bf 3}, 1067 (1971). 

\bibitem{LozYud}
Yu. E. Lozovik and V. I. Yudson, {\it On the ground state of the two-dimensional non-ideal bose gas}, Physica (Amsterdam) {\bf 93}A, 493 (1978).

\bibitem{kosterlitz}
J. M. Kosterlitz, {\it The critical properties of the two-dimensional XY model}, J. Phys. C {\bf 7}, 1046 (1974).

\end{thebibliography}
 \end{document}